\documentclass[10pt]{article}

\usepackage{lmodern}
\usepackage{amssymb,amsmath}
\usepackage{amsfonts}
\usepackage{algorithm}
\usepackage{algpseudocode}
\usepackage{natbib}

\usepackage{calrsfs}
\DeclareMathAlphabet{\pazocal}{OMS}{zplm}{m}{n}

\usepackage{geometry}
\usepackage{fancyhdr}
\usepackage{lastpage}

\let\emph\textit
\usepackage{graphicx}
\usepackage{longtable}
\usepackage{color}
\usepackage{hyperref}
\usepackage{booktabs}
\usepackage{tikz}
\usepackage{pgf}  
\usepackage{subcaption} 
\graphicspath{{FIG/}{SHARE/}{./}}
\newcommand{\vect}[1]{\boldsymbol{#1}}

\definecolor{refcol}{rgb}{0.5,0.5,0.5}

\definecolor{myblue}{rgb}{0.1,0.1,0.6}
\usepackage{cleveref}

\author{Mathias Lamari$^a$, Pierre Kerfriden$^a$, Oguz Umut Salman$^b$, \\Vladislav Yastrebov$^a$, Kais Ammar$^a$, Samuel Forest$^a$}
\title{A time-discontinuous elasto-plasticity formalism to simulate instantaneous plastic flow bursts}
\newcommand{\shortauthor}{M. Lamari et al.}
\newcommand{\shorttitle}{Time-discontinuous plasticity}
\date{\small$^a$\textit{Mines Paris, PSL University, Centre des matériaux, CNRS UMR 7633, BP 87, 91003, Evry, France}\\ $^b$\textit{LSPM, University Sorbonne Paris Nord, 93430, Villetaneuse, France} \\ [1em]\today}

\hypersetup{
  pdfauthor={Author},
  pdftitle={Title},
  colorlinks=true,
  linkcolor=myblue,
  filecolor=magenta,      
  urlcolor=myblue,
  citecolor=myblue,
}

\newcommand{\tens}[1]{\boldsymbol{#1}}

\newcommand{\tenstwo}[1]{\boldsymbol{#1}}
\newcommand{\tensfour}[1]{\mathbb{#1}}

\newcommand{\discont}[1]{[\![#1]\!] }

\usepackage{stmaryrd}
\usepackage{algorithm}
\usepackage{algpseudocode}

\begin{document}
\maketitle

\begin{abstract}
  Plastic flow is conventionally treated as continuous in finite element (FE) codes, whether in isotropic, anisotropic plasticity, or crystal plasticity. This approach, derived from continuum mechanics, contradicts the intermittent nature of plasticity at the elementary scale. Understanding crystal plasticity at micro-scale opens the door to new engineering applications, such as microscale machining.
In this work, a new approach is proposed to account for the intermittence of plastic deformation while remaining within the framework of continuum mechanics. We introduce a material parameter, the plastic deformation threshold, denoted as $\Delta p_{min}$, corresponding to the plastic deformation carried by the minimal plastic deformation burst within the material. The incremental model is based on the traditional predictor-corrector algorithm to calculate the elastoplastic behavior of a material subjected to any external loading. The model is presented within the framework of small deformations for von Mises plasticity. To highlight the main features of the approach, the plastic strain increment is calculated using normality rule and consistency conditions, and is accepted only if it exceeds $\Delta p_{min}$. To achieve this, a time-discontinuous generalization of the Karush-Kuhn-Tucker (KKT) conditions is proposed. The simulations show that the introduction of the plastic threshold allows for the reproduction of the spatiotemporal intermittence of plastic flow, capturing the self-organization of plastic flow in complex loading scenarios within an FE model.
\end{abstract}


\vspace{1em}
\noindent\textbf{Supplementary material:} The source code is available in the following \href{https://zenodo.org/records/14266823}{zenodo repository}.
\normalsize
  
\tableofcontents

\section{Introduction}

Plastic flow is usually modeled as a continuous spatio-temporal deformation mechanism in engineering finite element (FE) codes, be it in the context of isotropic, anisotropic, or crystal plasticity. This assumption underlying continuum plasticity models contradicts the intermittent nature of plasticity at elementary scales \citep{Brown2012}. This intermittency has been experimentally observed in various ways. For instance, during the compression of micropillars, the stress-strain curve exhibits significant serrations, which is related to the abrupt activation of a small number of slip planes~\citep{Zhang2017,Uchic2009}. Similar results are observed during the tension of submicrometer single crystals~\citep{Oh2009,Kiener2008,Borasi2023}. Acoustic emission and high-resolution extensometry experiments on centimeter-sized single-crystal samples also reveal the organized, non-chaotic nature of plastic flow due to dislocation avalanches and their interactions~\citep{Weiss2007}. Digital image correlation (DIC) at microscopic scale also demonstrates the presence of localized deformation bands at grain scale in macro-size polycrystalline materials  due to local hardening (nano-precipitates) followed by softening~\citep{Charpagne2021,Marano2019}. When the tested specimen is macroscopic, or the material contains a high density of grain boundaries, dislocation avalanches are hindered, and the flow becomes quasi-continuous, explaining why it has not been necessary to model this intermittency in traditional macroscopic industrial problems. However, with advances in nanotechnology and the increasing production of micro-metric components, developing a model that accounts for the spatial and temporal intermittency of plasticity and is usable in FE codes becomes necessary~\citep{ZepedaRuiz2021}. Understanding crystalline plasticity at this scale opens the door to a wide range of new engineering applications, such as micro-machining and the design of novel materials like hierarchical steels~\citep{XuZhao2022,Zhu2016,Xu2022}.

Let us acknowledge that at atomic scales, molecular dynamics (MD)~\citep{Castellanos2021,Patinet2011} and discrete dislocation dynamics (DDD)~\citep{Gomez-Garcia2006,Csikor2007} models accurately and "naturally" reproduce the correlated nature of dislocation avalanches and their power law amplitude distribution~\citep{Weiss2007}. While these models help understand the physics behind the intermittency and organization of plastic flow, they cannot be used at industrial scales due to excessively high computation times.

Several microscopic plasticity models have been recently proposed in the literature to account for the statistical structure of intermittent acoustic emission generated by plastic flow bursts without deviating (too much) from the continuum mechanics formalism. A noteworthy example of such models is the mesoscale tensorial model (MTM)~\citep{Salman2011,Salman2012,Baggio2023,Baggio2023bis}. Models that attempt to introduce intermittency into crystal plasticity also exist, typically relying on the introduction of stochasticity in an internal system variable (dislocation density~\citep{Weiss2015}, additional stress added to the yield limit~\citep{Wijnen2021,Vermeij2024}, critical resolved shear stress~\citep{Gelebart2021}) or within the solver itself~\citep{Yu2021,YuMarian2021}. These modifications succeed in predicting the intense localization of plastic deformation into finite bands, and size dependence of the yield stress in microscopic specimens.  However, they rely on introducing probabilities into the crystalline plasticity model itself and thus require prior knowledge of the laws governing the correlations of plastic flows. They are also computationally expensive due to the stochastic treatment at each time step. Recently, Ruestes and Segurado~\citep{Ruestes2024} used a kinetic Monte Carlo selection process to control the activation of microscopic slip mechanisms, achieving similar results. Without introducing any stochasticity, Ry{\'{s}} et al~\citep{Rys2024} have demonstrated that spontaneous localization of plastic strain can occur in crystal plasticity due to the yield-vertex effect when the latent-hardening of slip systems is higher than the self-hardening.

In this work, we propose a new approach to account for the intermittency of plastic deformation that does not rely on the prior knowledge of the statistics of plastic flow bursts. We start from a standard isotropic rate-independent J2-plasticity formulation, which is implemented in a finite element solver using the standard implicit time-stepping with radial return algorithm. We introduce a single strictly positive parameter, the plastic strain threshold denoted as $\Delta p_\text{min}$, which corresponds to the smallest (cumulative) plastic strain increment that the material may undergo at a particular time under load. This plastic strain threshold is physically associated to the smallest dislocation avalanche that can develop. Solving the equation of elasto-plasticity under the constraint that the plastic increment cannot be smaller than the plastic threshold is performed with very limited modification of the standard numerical formulation of departure. Typically, our implementation in the \href{https://docs.fenicsproject.org/}{FEniCSX} FEA library~\citep{Scroggs2022} requires modifying a single line of code to take this inequality constraint into account and depart from standard continuum finite element plasticity. Although the proposed model was derived by effectively manipulating a standard continuum plasticity finite element code to trigger the intermittence effects, we propose a set of constitutive equations for the time-discontinuous plasticity model, of which the modified FEA model mentioned previously is a rigorous implicit time-stepping implementation. Our presentation of the method will then follow a rather classical mode of exposure, starting from the continuum equations before deriving the associated fully discrete finite element model. A validation of the approach is made by implementing the model in another finite element code, namely \href{http://www.zset-software.com/}{Zset}~\citep{Besson1997}, thus showing that the model goes beyond a particular implementation.

The simulations performed using this modified, discontinuous, engineering plasticity formulation are remarquable in several ways. Noteworthily, the plastic flow arises in the form of plastic bands, which results in serrations on the load-displacement curve of tensile specimens. These bands are akin to the one observed in Portevin-Le-Chatelier (PLC) simulations~\citep{Colas2014,WangMaziere2011,Lamari2024,Guillermin2023}, but in contrast to strain ageing effects, they result from a rate-independent model formulation. In this sense, the phenomenology of the plastic flow generated by the proposed formulation shares resemblance with the Lüders effect, that would in our case be triggered repeatedly rather than once in the case of Lüders band propagation. Remarkably, the proposed model is discontinuous but does not exhibit features of instability encountered in softening plasticity models. There is no dependency of the band statistics (intensity and width) to mesh size (providing that the level of refinement is sufficient to describe strain gradients). In addition, it will be shown that the Newton solver associated with the finite element force balance equations converges in a well-behaved manner, exhibiting the expected quadratic local convergence.

The paper is organized as follows. In section 2, the physical motivation of the proposed model is given, followed by a presentation of the time-discontinuous model and its FE discretization. In section 3, a first simulation under uniform fields is presented, followed by a reference simulation with complex dogbone geometry. The methodology to analyze and quantify the complex plastic activity appearing in the simulations is then given. Section 4 is dedicated to the use of this methodology to analyze the effect of different simulations and model parameters on the model response. Finally, section 5 is devoted to the analysis of power law distributed plastic events, spatio-temporal analysis of plastic avalanches. Finally, the results of a tension test of a plate with a hole are discussed to highlight the potential of the model for application to structural computations.

\section{Model formulation}

\subsection{Physical foundation}

Recently, Perchikov and Truskinovsky have emphasized the fundamentally quantized nature of plastic deformation in crystalline materials \citep{Perchikov2024}, which occurs in a time-independent manner. In most metallic crystalline materials at room temperature, the most common deformation mechanism is dislocation slip. The displacement of a dislocation line induces a local displacement of $b$, which is the norm of the Burgers vector (typically $\approx 2 \times 10^{-10}$~m in iron \citep{Lamari2024}). In micro-scale samples (sample size $L \approx 10^{-6}$~m) the displacement of a dislocation from one Peierls valley to the next induces a plastic deformation of $\approx 2 \times 10^{-4}$, which is small but not infinitesimal. In situ observation of dislocation gliding shows a phenomenon that is very abrupt, occurring at the speed of sound in metals \cite{Oh2009,Legros2015}. From the point of view of the observer it can be seen as instantaneous. Therefore, in this work we will present a model that is time-discontinuous, time-independent and with a minimum plastic strain increment, called the plastic threshold and denoted $\Delta p_\text{min}$ with a value of $\approx 2 \times 10^{-4}$.

In single crystals, the absence of grain boundary screening of the dislocation long-range stress makes the displacement of one dislocation prone to induce motion of other dislocations in parallel slip planes, creating an avalanche of plastic deformation~\cite{Brown2012}. In micro-scale materials, dislocation sources are rare and hindered by surface effects~\citep{Zhang2017,Uchic2009}. This makes the system prone to reach self-organised criticality, leading to larger dislocation avalanches and scale-free intermittent plasticity~\citep{Dimiduk2006}. Therefore, the model we propose must be scale-free and prone to promote plastic strain bursts (i.e. plastic deformation increment > $\Delta p_\text{min}$). 

While the model is primarily intended for crystal plasticity to be comparable with metallic crystal experiments, for simplicity we will propose our model in the framework of isotropic J2 plasticity. This is not completely unphysical, since intermittent plasticity is also found in metallic glasses~\citep{Okuyucu2023}. While dislocation gliding is not the mechanism of plastic deformation in metallic glasses, the various mechanisms~\citep{Greer2013} such as the shear transformation zone (STZ)~\citep{Chevalier2018,Yan2010} also induce plastic deformation that is not locally infinitesimal. Plastic bands are also observed. The plastic threshold concept still applies and we will take its value to be $\approx 2 \times 10^{-4}$ for simplicity. 

{\color{black}In the following, the terms plastic burst or plastic jump are primarily used to describe significant plastic events occurring over a substantial region of the specimen. The term avalanche is used occasionally, in line with its usage in the physics community, despite the fact that the simulation does not directly involve dislocation concepts and that the plastic burst occurs in only one numerical step, as will be shown below.}

\subsection{Continuum plasticity}

In this section, the equations of basic continuum elasto-plasticity are recalled for a classical rate-independent J2-plasticity including an isotropic hardening law. {\color{black} This will clarify where the time-discontinuous elasto-plasticity model introduced in the following section diverges from the classical continuous framework.}

\paragraph{\textbf{Linearised continuum mechanics}}

The equations of elasto-plasticity {\color{black}are solved} over a space-time domain $\Omega \times \mathcal{T}$ where $\Omega \subset \mathbb{R}^3$ and $\mathcal{T}=[0, \ T]$. Let $\vect{x} \in  \mathbb{R}^3$ denote the spatial coordinates of a point of $\Omega$ and let $t \in \mathcal{T}$ denote the time variable. The boundary $\partial \Omega$ is split into a part $\partial_\mathrm{u}  \Omega$ over which time-dependent Dirichlet conditions are applied, and a part $\partial_\mathrm{t} \Omega$ over which homogeneous Neumann conditions are applied.

The displacement $ (\vect{x},t)  \mapsto \vect{u}(\vect{x},t)$ and Cauchy stress tensor field $ (\vect{x},t)  \mapsto  \tenstwo{\sigma} (\vect{x},t)  $ {\color{black}satisfies} the following equations:
\begin{equation}
\textrm{div} \, \tenstwo{\sigma}  = \vect{0} \qquad \textrm{in} \ \Omega
\end{equation}
\begin{equation}
\vect{u} = \vect{u}_d \qquad \textrm{over} \ \partial_\mathrm{u}  \Omega
\end{equation}
\begin{equation}
 \tenstwo{\sigma}  \cdot \vect{n}_{\partial \Omega} = \vect{0} \qquad \textrm{over} \ \partial_\mathrm{t}  \Omega
\end{equation}
where $\vect{n}_{\partial \Omega} $ denotes the outer normal to the domain boundary and $\vect{u}_d$ is a known function defined over $\partial_\mathrm{u} \Omega \times \mathcal{T}$ and with values in $ \mathbb{R}^3$. Volume forces, including inertial forces are excluded for simplicity.

The infinitesimal strain tensor is defined as
\begin{equation}
\tenstwo{\varepsilon} = \frac{1}{2} \left( {\nabla} \vect{u} +  \left({\nabla} \vect{u}\right)^T  \right)
\end{equation}

To close the system, a constitutive relation linking $\tenstwo{\sigma} $ to the time-history of $\tenstwo{\varepsilon}$ needs to be introduced.

\paragraph{\textbf{Constitutive equations}}
Two additional unknown fields of internal variables are introduced, namely the second-order plastic strain tensor $ (\vect{x},t) \mapsto  \tenstwo{\varepsilon}^\mathrm{p}    (\vect{x},t)$ and the scalar cumulative plastic strain $ (\vect{x},t) \mapsto  p   (\vect{x},t)$. The constitutive relation of the elasto-plastic material is written as follows. 
The strain tensor can be decomposed into elastic and plastic parts
\begin{equation}
\tenstwo{\varepsilon} = \tenstwo{\varepsilon}^{\mathrm{e}} + \tenstwo{\varepsilon}^{\mathrm{p}}
\end{equation}
The Cauchy stress is given by Hooke's law for isotropic elasticity
\begin{equation}
\label{eq:elast}
 \tenstwo{\sigma} = \tensfour{C} : ( \tenstwo{\varepsilon} -  \tenstwo{\varepsilon}^\mathrm{p} ) = \lambda \, \mathrm{tr}(\tenstwo{\varepsilon}^{\mathrm{e}}) \tenstwo I + 2\mu \, \tenstwo{\varepsilon}^{\mathrm{e}}
\end{equation}
where $\tensfour{C} $ is the fourth-order elasticity tensor, $\lambda,\mu$ are Lam\'e coefficients and $\mathrm{tr}(\tenstwo{\varepsilon}^{\mathrm{e}})$ denotes the trace of tensor $\tenstwo{\varepsilon}^{\mathrm{e}}$. The yield surface is defined as
\begin{equation}
f( \tenstwo{\sigma};p) = \sigma_\mathrm{vM}( \tenstwo{\sigma} ) - R(p) - \sigma_\mathrm{ys} 
\end{equation}
where $R(p)$ is the isotropic hardening function which is assumed to be monotonically non-decreasing, $\sigma_\mathrm{y} $ is the initial yield stress and {\color{black}$\sigma_\mathrm{vM}$ the von Mises equivalent stress ($\sigma_\mathrm{vM} = \sqrt{ \frac{3}{2} \tenstwo{s}:\tenstwo{s}   }$ where $\tenstwo{s}$ is the deviatoric part of the Cauchy stress tensor)}. 
The normal to the yield surface is defined as $ \tenstwo{n} = \partial f / \partial \tenstwo{\sigma} $. The normality rule for the plastic flow is:
\begin{equation}
\label{eq:normality}
 \dot{\tenstwo{\varepsilon}}^\mathrm{p}  = \lambda \, \tenstwo{n} 
\end{equation}
The system is closed by the Karush-Kuhn-Tucker (KKT) conditions:
\begin{equation}
\label{eq:yield_cond}
f \lambda = 0
\end{equation}
\begin{equation}
\label{eq:positivityLambda}
\lambda \geq 0
\end{equation}
\begin{equation}
\label{eq:surf}
f \leq 0  
\end{equation}
The time-integrated multiplier coincides with the cumulative plastic strain $p$ in J2-plasticity. This specificity reads as:
\begin{equation}
\lambda = \dot{p} = \sqrt{\frac{2}{3}  \dot{\tenstwo{\varepsilon}}^\mathrm{p} :  \dot{\tenstwo{\varepsilon}}^\mathrm{p} }
\end{equation}

\paragraph{\textbf{Initial conditions}} 

Finally, initial conditions need to be applied to the fields of internal variables, which {\color{black}take} the form $p_{|t=0}=0$ and $  \tenstwo{\varepsilon}^\mathrm{p}_{|t=0}= \tenstwo{0}$.

\subsection{Time-discontinuous plasticity model}

In the time-discontinuous plasticity model, the cumulative plastic strain variable $p$ {\color{black}is allowed} to increase \emph{only} by instantaneous events (bursts) of finite magnitude. To do this, mechanical quantities are allowed to operate beyond the yield surface, i.e. violate the condition \eqref{eq:surf}, and return to the yield surface when and only when the resulting instantaneous increment of cumulative plastic strain is larger than $\Delta p_\mathrm{min}~\in~\mathbb{R}^{+*}$. The plastic threshold $\Delta p_\mathrm{min}$ is the sole additional parameter of the model, and represents the smallest plastic strain increment that the material system can accommodate.

To achieve this mathematically, Eq. \eqref{eq:normality}, \eqref{eq:yield_cond} and \eqref{eq:surf} are modified, {\color{black}while the others are retained}. The first equation is altered as follows
\begin{equation}
\label{eq:normality2}
 \llbracket  \tenstwo{\varepsilon}^\mathrm{p} \rrbracket =   \llbracket  p \rrbracket \, \tenstwo{n}^-
\end{equation} 
In this equation,  $ \llbracket \bullet \rrbracket =  \lim_{{h \to 0}} [\bullet(t+h) - \bullet(t-h)] $, with $\bullet$ being any quantity such as $p$ or $\tenstwo{\varepsilon}^\mathrm{p}$, indicates a jump of the considered quantity at time $t$. Notation $ \tenstwo{n}^- = \lim_{{h \to 0}} \tenstwo{n}(t-h)  $ denotes the normal to the yield surface at time $t$, taking the limit from the left if the normal evolves in a discontinuous manner in time, as this corresponds to the stress state before the {\color{black}plastic burst}\footnote{Notice that in J2-plasticity, the normal tensor remains unchanged during a plastic burst, i.e.
\begin{equation}
\nonumber
\label{eq:normalInvariance}
 \llbracket \tenstwo{n}\rrbracket = 0
\end{equation} 
}. Physically, this means that the stress state before the plastic burst occurs determines the direction of plastic yield.

{\color{black} At each time, the plastic strain increment must satisfy new constitutive equations. The set $\mathcal{P}^{(t)}$  is defined as the collection of admissible plastic bursts that comply with the constitutive relations presented herein. Admissible cumulative plastic strain increments that belong to the set $\mathcal{P}^{(t)}$ will be denoted as $\llbracket p \rrbracket^\star $. The corresponding evolution of yield surface after the increment $\llbracket p \rrbracket^\star $ is denoted ${f^+}^\star$, defined as ${f^+}^\star = f( \tensfour{C} : ( \tenstwo{\varepsilon} -  \tenstwo{\varepsilon}^\mathrm{p} -\llbracket p \rrbracket^\star \tenstwo{n}^-) , p^- + \llbracket p \rrbracket^\star    )$, with $p^-  = \lim_{{h \to 0}} p(t-h)$.} 

{\color{black}The first new equation of the time-discontinuous plasticity model that controls the value of $\llbracket p \rrbracket^\star $ is:}
\begin{equation}
\label{eq:yield_cond2}
\llbracket p \rrbracket^\star  {f^+}^\star = 0
\end{equation}
Eq. \eqref{eq:yield_cond2} is used to express the fact that if there is a non-vanishing instantaneous plastic flow increment at time $t$, its amplitude must be such that the mechanical stress returns to the yield surface.

{\color{black}Finally, the plastic strain increment must obey a last equation assuring that the system is closed:}
\begin{equation}
\label{eq:surf2}
\llbracket p \rrbracket^\star \left<  \Delta p_\mathrm{min} -  \llbracket p \rrbracket^\star  \right> = 0
\end{equation}
where $  \left<  \, . \, \right> = \max \{ 0,   \, . \,  \}$. This equation means that either the plastic flow is null, or the increment of cumulative plastic strain is larger than $\Delta p_\mathrm{min}$, which introduces flow discontinuities in the model.

Then, at time $t \in \mathcal{T}$, the jump of cumulative plastic strain is chosen as the maximum of all admissible cumulative plastic strains:
\begin{equation}
\label{eq:max}
\llbracket p \rrbracket = \underset{ \llbracket p \rrbracket^\star \in\mathcal{P}^{(t)}  }{ \max  } \llbracket p \rrbracket^\star
\end{equation}
Indeed, the set of Eq. \eqref{eq:yield_cond2} and \eqref{eq:surf2} introduced previously leaves an indeterminacy. This is because $\llbracket p \rrbracket=0$ is always a trivial solution. The max operation in Eq. \eqref{eq:max} specifies that if a jump in plastic strain is admissible and positive, then it must happen.

{\color{black} The term "time-discontinuous" refers to the fact that when a plastic burst occurs, certain quantities experience a discontinuity in their temporal evolution, such as $p$, $f$,  $\tenstwo{\varepsilon_\mathrm{p}}$ due to Eq. \eqref{eq:yield_cond2} and \eqref{eq:surf2}. However, this is not the case for all quantities, such as $\tenstwo{\sigma}$, $\tenstwo{\varepsilon_\mathrm{e}}$, $\tenstwo{\varepsilon}$ or $\tenstwo{n}$ which may exhibit either time-continuous or discontinuous behavior depending on the external conditions.}

\subsection{Algorithmic implementation} \label{sec:ConstitutiveUpdate}

{\color{black} This section outlines the algorithm used to implement the time-discontinuous model. A straightforward modification to the radial return algorithm ensures that plastic strain follows the new constitutive equations from Eq. \eqref{eq:normality2} to \eqref{eq:max}, with the remainder of the procedure remaining classical. A comprehensive description of the model's implementation in the finite element solver, the time discretization of all constitutive relations, and the integration of the global Newton solver is provided in Appendix A.}

{\color{black} The time interval $\mathcal{T}=[0 \ T]$ is divided into $N$ time steps. The equation of continuum mechanics {\color{black}is enforced} at discrete times $\bar{\mathcal{T}} = \{ t_0 , t_1 , \, ... \, , t_{N} \}$, using the finite element method. In the following, $n$ will be an integer between 0 and $N$. As is customary, the quantity $X$ at step $n$ will be denoted $X_n$.}

{\color{black} Suppose the system's evolution is known up to step $n$.} The output of stress function $\tenstwo{\sigma}_{n+1}(  \tenstwo{\varepsilon}_{n+1} ;  \tenstwo{\varepsilon}^\mathrm{p}_{n}  , p_{n} )$ is evaluated via an operator-splitting procedure akin to those that are  traditionally employed in implicit schemes for classical rate-independent elasto-plastic constitutive laws~\citep{Besson2009,Souza2011,Simo1998}. This is done as follows, at time $t_{n+1}$:

\begin{itemize}
\item {\color{black}Determine the direction of the plastic flow
 \begin{equation}
\tenstwo{n}^-_{n+1} = \left. \frac{\partial f }{\partial \sigma} \right|_{ \tenstwo{\sigma}_{n+1} = \tensfour{C} : ( \tenstwo{\varepsilon}_{n+1} -  \tenstwo{\varepsilon}^\mathrm{p}_{n} ) , \, p_{n+1} = p_n} 
\end{equation}}
\item Evaluate yield function $f^\star$ assuming that no plastic increment occurs. 
 \begin{equation}
f^\star_{n+1} = \sigma_\mathrm{eq} \left( \tensfour{C} : ( \tenstwo{\varepsilon}_{n+1}   -   \tenstwo{\varepsilon}^\mathrm{p}_{n}  )  \right) - R(p_{n} ) - \sigma_\mathrm{y}
\end{equation}
\item {\color{black}Discriminate between elastic and plastic behavior:}
\begin{itemize}
\item if $f^\star_{n+1} <=0$, set  $p_{n+1} = p_n$, 
\item if $f^\star_{n+1} >0$, perform the following two steps:
\begin{itemize}
\item assume that a plastic jump occurs within time step $]t_{n} \  t_{n+1}]$, i.e. $ (p_{n+1} - p_n) \neq 0$ (in that case, Eq. \eqref{eq:yield_cond2} specifies that the stress must return to the yield limit at the end of the time step). Look for test plastic increment $\Delta p^\star$ satisfying
\begin{equation}
 \sigma_\mathrm{eq} \left( \tensfour{C} : ( \tenstwo{\varepsilon}_{n+1}   -   \tenstwo{\varepsilon}^\mathrm{p}_{n}  -  \Delta p^\star \tenstwo{n}^-_{n+1} )  \right) - R(p_{n} + \Delta p^\star) - \sigma_\mathrm{ys} = 0
\end{equation}
using a Newton-Raphson algorithm.
\item Compare $\Delta p^\star$ to $\Delta p_\mathrm{min} $. If $\Delta p^\star < \Delta p_\mathrm{min} $, set $p_{n+1} = p_n$, otherwise set $p_{n+1} = p_n + \Delta p^\star$
\end{itemize}
\end{itemize}
\item Compute $  \tenstwo{\varepsilon}^\mathrm{p}_{n+1} =  \tenstwo{\varepsilon}^\mathrm{p}_{n} +   ( p_{n+1} -p_{n}  ) \tenstwo{n}^-_{n+1} $ 
and return $\sigma_{n+1} =  \tensfour{C} : \left( \tenstwo{\varepsilon}_{n+1}  - \tenstwo{\varepsilon}^\mathrm{p}_{n+1}  \right) $
\end{itemize}

A pseudo-code of the radial return algorithm as implemented in the \href{https://docs.fenicsproject.org/}{FEniCSX} code is given in Algorithm~\ref{Algorithm1}. Compared to a classical radial return algorithm used in continuum plasticity, only lines 14 to 17 have been added. Removing these lines will revert to the classical plasticity approach. The full script used in our FEniCSX simulation, including geometry and example runs, can be found in this \href{https://github.com/Mathias-Lamari/Time-discontinuous-plasticity}{GitHub repository}. Due to limitations related to automatic differentiation in FEniCSX, which does not support loops, the maximum number of iterations in the Newton loops, $N_\text{Max}$, has been set to 1. However, comparisons with Zset simulations show that this does not significantly affect the simulation results, even when the work-hardening behavior is not purely linear.

\begin{algorithm}
\caption{Time-discontinous radial return (von Mises)}
\label{Algorithm1}
\begin{algorithmic}[1]
	\State \textbf{Initialize:}  $\tenstwo{\varepsilon}, \tenstwo{\varepsilon}^\mathrm{p}, p, \tensfour{C}, \Delta p_{\text{min}}, \mu, R, \delta, N_\text{max} $
    \State \textbf{Calculate:} $\tenstwo{\sigma^\star} = \tensfour{C} : ( \tenstwo{\varepsilon}   -   \tenstwo{\varepsilon}^\mathrm{p})$
    \State \textbf{Calculate:} $f = \sigma_\text{vM} \left( \sigma^\star \right) - \sigma_\mathrm{ys}   - R(p) $
   	\State \textbf{Calculate:} $\tenstwo{n}^- = \partial f / \partial \sigma^\star  $
    \State \textbf{Set:} $i = 0$
    \State \textbf{Set:} $\Delta p^\star = 0$
    \While{$|f| \geq \delta$ and $i \leq N_\text{max}$}
    	\State \textbf{Calculate:} $f' = -3 \mu - \frac{dR}{dp} (p + \Delta p^\star)$
	    \State \textbf{Calculate:} $\Delta p^\star = - f / f' $
	    \State \textbf{Calculate:} $\tenstwo{\sigma^\star} = \tensfour{C} : ( \tenstwo{\varepsilon}   -   (\tenstwo{\varepsilon}^\mathrm{p}+\Delta p \cdot \tenstwo{n}^-))$
    	\State \textbf{Calculate:} $f = \sigma_\text{vM} \left( \sigma^\star \right) - \sigma_\mathrm{ys}   - R(p + \Delta p^\star) $
	    \State \textbf{Calculate:} $i = i+1$
	\EndWhile
    \If{$\Delta p^\star \geq \Delta p_{\text{min}}$}
        \State $\Delta p = \Delta p^\star $
    \Else
        \State $\Delta p = 0$
    \EndIf
    \If{$f \geq 0$}
        \State $\Delta \tenstwo{\varepsilon}^\mathrm{p} = \Delta p \cdot \tenstwo{n}^-$
    \Else
        \State $\Delta \tenstwo{\varepsilon}^\mathrm{p} = 0 \cdot \tenstwo{n}^-$
    \EndIf
    \State \textbf{Update:} $\tenstwo{\varepsilon}^\mathrm{p} = \tenstwo{\varepsilon}^\mathrm{p} + \Delta \tenstwo{\varepsilon}^\mathrm{p}$
    \State \textbf{Update:} $p = p + \sqrt{\frac{2}{3} \Delta \tenstwo{\varepsilon}^\mathrm{p}:\Delta \tenstwo{\varepsilon}^\mathrm{p}}$
    \State \textbf{Return:} $\tenstwo{\varepsilon}^\mathrm{p}, p$
\end{algorithmic}
\end{algorithm}

\subsection{Equivalence with a two-yield surface model}

{\color{black}This section demonstrates the equivalence of the time-discontinuous plasticity model to a two-yield surface model. Note that all new interpretations presented here are not additional properties of the model but are instead derived from previous equations.

It is assumed that, at time $t$,} a plastic burst occurs (i.e. $\discont p > 0$) {\color{black}in a region of the system}. The tensors $\tens{e}$ and $\tens{e}^e$ are the deviatoric parts of the total and elastic strain tensors ($\tens{\varepsilon}^p$ is always deviatoric in von Mises plasticity). Hooke's law provides the deviatoric stress component at the end of the plastic burst:
\begin{equation}
\tens{s} + \discont{\tens{s}} = \tens{s}+  2 \mu \discont{\tens{e}^e} = \tens{s} + 2 \mu (\discont{\tens{e}} - \discont{\tens{\varepsilon}^p})  =  \tens{s}^{\text{\tiny trial}} - 2 \mu \discont{\tens{\varepsilon}^p}
\label{Increm_Contr_Dev}
\end{equation}

\begin{figure}[!htb]
\centering
\includegraphics[width=0.6\textwidth]{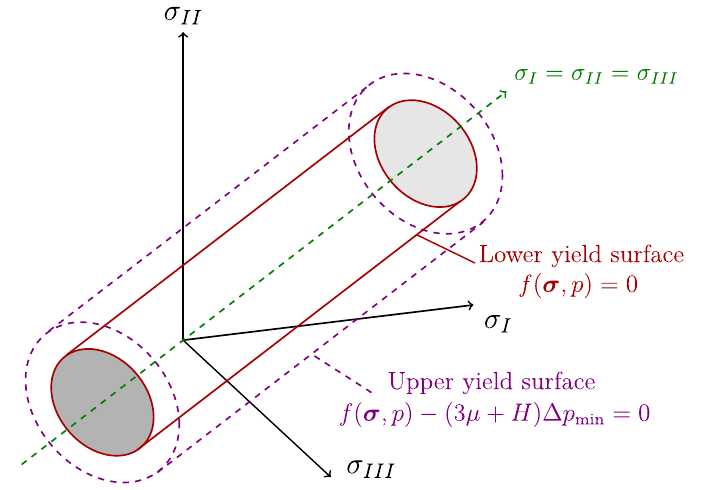}
\caption{Representation of the lower and upper yield surfaces in the principal stress basis. Yielding occurs when a material point exceeds the upper yield surface, and the plastic strain increment is determined relative to the lower yield surface.}
\label{fig:ThickYieldSurface}
\end{figure}

In Equation~\eqref{Increm_Contr_Dev}, the trial stress  $\tens{s}^{\text{\tiny trial}} = \tens{s} + 2 \mu \discont{\tens{e}}$  is {\color{black}the stress reached if all strain increment is elastic. The von Mises stress corresponding to $\tens{s}^{\text{\tiny trial}}$ is denoted $\sigma_{\text{\tiny vM}}^{\text{\tiny trial}}$.} From Equation~\eqref{Increm_Contr_Dev}, the tensor $\tens{s}^{\text{\tiny trial}}$ is the linear combination of $\tens{s}+\discont{\tens{s}}$ and $\discont{\tens{\varepsilon}^p}$. From Equation~\eqref{eq:normality2}, $\discont{\tens{\varepsilon}^p}$ is colinear to $\tens{s}+\discont{\tens{s}}$, {\color{black} since} the normal to the yield surface $\tens n$ is the same before and after plastic relaxation). Then $\tens{s}^{\text{\tiny trial}}$ is colinear to $\tens{s}+\discont{\tens{s}}$. Therefore, the following normalized tensors are equal:
\begin{equation}
\sqrt{\frac{3}{2}} \frac{\tens{s}+\discont{\tens{s}}}{\sigma^+_{\text{vM}}} = \sqrt{\frac{3}{2}} \frac{\tens{s}^{\text{\tiny trial}}}{\sigma_{\text{ vM}}^{\text{\tiny trial}}} \quad \text{where} \quad \sigma^+_\text{ vM} = \sqrt{\frac{3}{2} (\tens{s}+\discont{\tens{s}}) : (\tens{s}+\discont{\tens{s}})}
\label{TenseurNormaliseEgalite}
\end{equation}
We deduce from the last {\color{black}Eq. \eqref{Increm_Contr_Dev} and \eqref{TenseurNormaliseEgalite}, and the normality rule of plastic flow Eq. \eqref{eq:normality2} } that:
\begin{equation}
\sigma^+_\text{ vM}  =  \sigma_{\text{\tiny vM}}^{\text{\tiny trial}} - 3 \mu \discont p
\label{Increm_Contr_vM}
\end{equation}
After {\color{black} the plastic jump}, the yield surface function is null:
\begin{equation}
\sigma^+_\text{ vM} = \sigma_{\text{ys}} + R(p+\discont p)
\label{SurfaceDeChargeInferieure}
\end{equation}
{\color{black} Combining Eq.~\eqref{Increm_Contr_vM} and \eqref{SurfaceDeChargeInferieure}, the relationship between $\sigma_{\text{\tiny vM}}^{\text{\tiny trial}}$ and $\discont p$ becomes straightforward:
\begin{equation}
\sigma_{\text{ vM}}^{\text{\tiny trial}} = \sigma_{\text{ys}} + R(p+\discont p) + 3 \mu \discont p
\label{EquationX}
\end{equation}}
By substituting the condition $\discont p  \geq \Delta p_{\text{min}} $ in Eq.~\eqref{EquationX}, {\color{black} the following condition on the von Mises stress is derived to predict the occurrence of plastic burst}:
\begin{equation}
\sigma_{\text{ vM}}^{\text{\tiny trial}} \geq \sigma_{\text{ys}} + R(p+\Delta p_{\text{min}}) + 3 \mu  \Delta p_{\text{min}}
\label{SurfaceDeChargeSuperieure}
\end{equation}
The latter condition is equivalent to defining an upper yield surface that describes the onset of yielding. {\color{black} Linear isotropic hardening is assumed for the remainder of the article to simplify the interpretation of simulations, i.e. $R(p) = H p$, $H$ being the hardening modulus. Note that the model can be applied to any form of non-decreasing hardening.}. Eq.~\eqref{SurfaceDeChargeSuperieure} is then simplified: 
\begin{equation}
\sigma_{\text{ vM}}^{\text{\tiny trial}} \geq \sigma_{\text{ys}} + H \ p + (3 \mu +H)  \Delta p_{\text{min}}
\label{SurfaceDeChargeSuperieure}
\end{equation}
It means that as soon as $\sigma_{\text{ vM}}^{\text{\tiny trial}}$ reaches this critical level, a plastic burst takes place.  In Fig. \ref{fig:ThickYieldSurface}, a representation of the lower and upper yield surfaces in the principal stress space is given. Yielding occurs when the stress in a material point of the system exceeds the upper yield surface, and the plastic strain increment is determined relative to the lower yield surface. The difference between the two yield surfaces is constant and equal to $(3 \mu +H)  \Delta p_{\text{min}}$. This constant gap explains visually why in the time-discontinuous model, the plastic strain increment cannot be infinitesimal for a given $\Delta p_{\text{min}}$.

\section{Results}

To explore qualitatively and quantitatively the properties of our time-discontinuous model, we first simulate a simple 1D geometry under uniaxial tension. Then, an analytical solution is found for homogeneous deformation under any triaxiality condition and compared with simulations. Finally, more complex geometries are tested under uniaxial tension. The model has been implemented independently in the \href{https://docs.fenicsproject.org/}{FEniCSX}~\citep{Scroggs2022} and \href{http://www.zset-software.com/}{Zset}~\citep{Besson1997} solvers and both solvers deliver the same results.

\subsection{Simulations with uniform fields}\label{sec:1DTensile}

\begin{figure}[!htb]
\centering
\includegraphics[width=\textwidth]{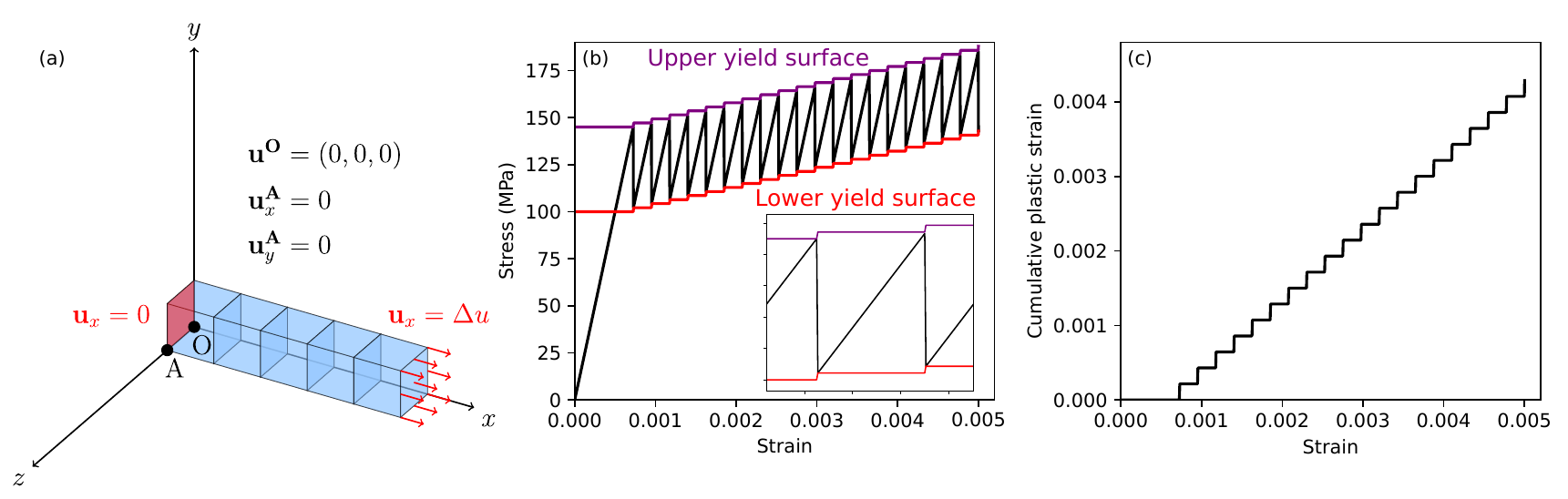}
\caption{(a) Geometry used for the homogeneous simulations. The number $N$ of aligned hexahedral elements can be set to any integer. Prescribed displacements are imposed on the left and right faces of the sample. Displacement components not indicated are left free. All other faces are free. (b) Evolution of stress $\sigma_{xx}$  as a function of strain $\varepsilon_{xx}$ for the homogeneous simulation. The lower yield surface (red) and the upper yield surface (purple) have been indicated. (c) Corresponding evolution of the cumulative plastic strain as a function of applied strain.}
\label{fig:HomogeneousResponse}
\end{figure}

A one-dimensional geometry consisting of N hexahedral elements aligned along the $O_x$ axis is used in this section, as depicted in Fig. \ref{fig:HomogeneousResponse} (a). On faces orthogonal to the $O_x$ axis, mixed conditions are imposed. At each numerical step, the displacement increment component along $O_x$ is fixed (0 for the left face at the origin, $\Delta u$ for the right face), while the other components are left free, except at points A and O, to suitably fix the rigid body motion (details in Fig. \ref{fig:HomogeneousResponse} (a)). The other surfaces are free. The initial yield stress $\sigma_{\text{ys}}$ is set to 100~MPa, the linear hardening rate $H$ to 10 GPa, the Young modulus $E$ to 200 GPa, and the Poisson ratio $\nu$ to 0.3. The imposed displacement step $\Delta u$ corresponds to a strain increment $\Delta \varepsilon_{xx}$ in the tensile direction of $3 \times 10^{-6}$ at each loading step. The plastic threshold $\Delta p_\text{min}$ is set to $2 \times 10^{-4}$. {\color{black} For the simulations to accurately reflect the time-discontinuous model under displacement boundary conditions, it is necessary to have, for all component i,j, that $\Delta \varepsilon_{ij} \ll \Delta p_\text{min}$.}

The first noteworthy feature of the simulation is the spatial homogeneity of both stress and strain fields across the entire geometry under tensile loading.  Without introducing defects in the mesh or the elastic/plastic properties of the material, there is no localization of stress, regardless of the number of elements. Under tensile condition, the stress tensor has only one nonzero component along the $O_x$ axis. This test is equivalent to testing a single Gauss point, but this setup was used in order to demonstrate the absence of localization in this particular case.

The tensile curve (stress $\sigma_{xx}$ vs strain $\varepsilon_{xx}$) is shown in Fig. \ref{fig:HomogeneousResponse} (b). The evolution of the lower (red) and upper (purple) yield surfaces as defined by Eq. \eqref{SurfaceDeChargeInferieure} and \eqref{SurfaceDeChargeSuperieure} are also incorporated for clarity, even though it must be emphasized that they are not used explicitly in the algorithm. The specimen first deforms elastically until the von Mises stress reaches the upper yield surface. {\color{black}Then, the entire specimen undergoes plastic deformation, causing the stress to decrease through elastic relaxation until it reaches the lower yield surface}. The plastic strain burst occurs within one numerical step. Then, the phenomenon repeats but with a hardened material due to linear isotropic hardening. Thus, periodic serrations are observed on the stress-strain curve; each stress increase corresponds to purely elastic loading of the material with a slope equal to the Young modulus; each drop corresponds to plastic deformation of the entire specimen.

Fig. \ref{fig:HomogeneousResponse} (c) gives the evolution of equivalent plastic strain during the simulation. As expected, plastic strain increases by steps, each plastic burst occurring in one numerical increment and corresponding to a stress drop on the tensile curve. {\color{black}Only two values of plastic strain increment $\discont p$ are possible at each numerical step, as demonstrated in the next section \ref{sec:HomogogeneousSection}, for any triaxiality conditions.} If no plastic deformation occurs, $\discont p$ is 0. If a {\color{black}plastic burst} occurs, $\discont p$ is given by:
\begin{equation}
\discont p = \frac{3 \mu + H}{E+H} \Delta p_\text{min}
\label{eq:dp}
\end{equation}
The last Eq. \eqref{eq:dp} is valid under uniaxial tension. When the Poisson ratio $\nu = 0.5$, then $\discont p = \Delta p_\text{min}$. The drop of von Mises stress at each plastic event $\discont {\sigma_\text{vM}}$ is given by:
\begin{equation}
\discont {\sigma_\text{vM}} = - E \discont p 
\label{eq:ds}
\end{equation}
The Eq. \eqref{eq:dp} and \eqref{eq:ds} are only valid for uniaxial straining.

\subsection{Analytical homogeneous solution under any triaxiality}\label{sec:HomogogeneousSection}

In this section, a single Gauss point is considered. The material parameters are chosen to be the same as in section \ref{sec:1DTensile}. We control the triaxiality factor $T$ defined as the ratio between the hydrostatic stress $\sigma_\text{h}$ and the equivalent stress $\sigma_\text{vM}$:
\begin{equation}
T = \frac{\sigma_\text{h}}{\sigma_\text{vM}} \quad \text{with} \quad \sigma_\text{h} = \text{tr} \left( \tenstwo{\sigma} \right)/3
\label{eq:triaxiality}
\end{equation}

For all tested $T$ values, which ranged from 0 (pure shear) to 2/3 (equibiaxial), the results in terms of equivalent stress and equivalent plastic strain are qualitatively identical to what is presented in Fig. \ref{fig:HomogeneousResponse}. Periodic stress drops linked with plastic {\color{black}bursts} are found, and both $\discont p$ and $\discont {\sigma_\text{vM}}$ admit only one given value during plastic bursts, as in the tensile case. For pure shear, Eq. \eqref{eq:dp} is not satisfied and we have instead $\discont p = \Delta p_\text{min}$, independently of the values of $E$ and $\nu$.

To understand this difference and calculate the theoretical value of the stress drop and plastic strain increase during a {\color{black}plastic burst}, we consider a time $t$ at which a plastic burst occurs. Independently of the triaxiality factor, the two-yield surfaces concept is respected. The value of the von Mises stress just before the jump corresponds to the value of the upper yield surface prior to the plastic strain increment, i.e.:
\begin{equation}
\sigma_\text{vM} = \sigma_\mathrm{ys} + H p + (3 \mu + H) \Delta p_\text{min}
\end{equation}
The value of the equivalent stress after plastic deformation corresponds to the value of the lower yield surface accounting for the plastic strain increment:
\begin{equation}
\sigma_\text{vM} + \discont{\sigma_\text{vM}}= \sigma_\mathrm{ys} + H (p + \discont p) 
\end{equation}
Therefore:
\begin{equation}
\discont{\sigma_\text{vM}}= H \discont p - (3 \mu + H) \Delta p_\text{min}
\label{eq:SvM_incrementHomogeneous}
\end{equation}

In the last Eq. \eqref{eq:SvM_incrementHomogeneous}, $\discont{\sigma_\text{vM}}$ and $\discont{p}$ are both unknown, and a second equation is necessary to obtain their values. This second equation is obtained by using the prescribed total deformation. {\color{black} As previously stated in section \ref{sec:1DTensile}, each component of the strain tensor at each numerical step must be negligible with regards to the plastic threshold $\Delta p_\text{min}$}. Therefore, even during the {\color{black}plastic burst}, the total strain tensor of the system does not change:
\begin{equation}
\discont{\tenstwo{\varepsilon}} = \discont{\tenstwo{\varepsilon}^{\mathrm{e}}} + \discont{\tenstwo{\varepsilon}^{\mathrm{p}}} = 0
\label{eq:NegligibleStrain}
\end{equation}

In continuum plasticity, if the strain rate is negligible, then the plastic strain rate and elastic strain rate are both individually negligible. One important difference in our time-discontinuous approach is that while $\discont {\tenstwo \varepsilon}$ is zero during the instantaneous {\color{black}plastic burst}, $\discont {\tenstwo{\varepsilon}^{\mathrm{e}}}$ and $\discont {\tenstwo{\varepsilon}^{\mathrm{p}}}$ are not infinitesimal, and we have:
\begin{equation}
\discont{\tenstwo{\varepsilon}^{\mathrm{e}}} = - \discont{\tenstwo{\varepsilon}^{\mathrm{p}}}
\label{eq:el_equals_pl_1}
\end{equation}

To obtain a scalar relationship between $\discont p$ and $\discont {\sigma_\text{vM}}$, the tensor contraction of both sides of Eq. \eqref{eq:el_equals_pl_1} with $\discont{\tenstwo \sigma}$ is taken. The right-hand side is simplified by using successively the normality rule (Eq. \eqref{eq:normality2}), the deviatoric nature of the normal tensor $\tenstwo{n}$ and the invariance of $\tenstwo{n}$ during the {\color{black}plastic burst}:
\begin{equation}
\discont{\tenstwo \sigma} : \discont{\tenstwo{\varepsilon}^{\mathrm{p}}} = \discont{p}  \left( \discont{\tenstwo \sigma} : \tenstwo{n} \right) = \discont{p} \left( \discont{\tenstwo s} : \tenstwo{n} \right) = \discont{p} \discont{\sigma_\text{vM}}
\label{eq:el_equals_pl_2}
\end{equation}

The tensor contraction $\discont{\tenstwo \sigma} : \discont{\tenstwo{\varepsilon}^{\mathrm{e}}}$ is simplified by decomposing the stress and elastic strain tensors into their hydrostatic part $\tenstwo{\sigma}_\text{h}$ and $\tenstwo{\varepsilon}_\text{h}^{\mathrm{e}}$ and their deviatoric part $\tenstwo s$ and $\tenstwo e^{\mathrm{e}}$, and using the Hooke law ($K$ is the bulk modulus):
\begin{equation}
\discont{\tenstwo \sigma} : \discont{\tenstwo {\varepsilon}^{\mathrm{e}}} = \discont{\tenstwo{\sigma}_\text{h}} : \discont{\tenstwo{\varepsilon}^{\mathrm{e}}_\text{h}} + \discont{\tenstwo{s}} : \discont{\tenstwo{e^{\mathrm{e}}}} = \frac{1}{3K} \discont{\tenstwo{\sigma}_\text{h}}^2 + \frac{1}{2 \mu} \discont{\tenstwo{s}}^2
\label{eq:el_equals_pl_3}
\end{equation}

The expression \ref{eq:el_equals_pl_3} is furthermore simplified using $\discont{\tenstwo{\sigma}_\text{h}}^2 = 3 \discont{\sigma_\text{h}}^2 = 3 T^2 \discont{\sigma_\text{vM}}^2$ and $ \discont{\tenstwo{s}}^2 = 2/3 \discont{\sigma_\text{vM}}^2$:
\begin{equation}
\discont{\tenstwo \sigma} : \discont{\tenstwo {\varepsilon}^{\mathrm{e}}} = \left( \frac{T^2}{K}  + \frac{1}{3 \mu}\right) \discont{\sigma_\text{vM}}^2
\label{eq:el_equals_pl_4}
\end{equation}

Combining Eq. \eqref{eq:el_equals_pl_2} and \eqref{eq:el_equals_pl_4}, a second relationship between  $\discont p$ and $\discont {\sigma_\text{vM}}$ is thus obtained:
\begin{equation}
\discont {\sigma_\text{vM}} = - M_T \discont p \quad \text{with} \quad M_T = \frac{1}{\frac{T^2}{K}  + \frac{1}{3 \mu}} = \frac{3 \mu E }{(27 \mu - 9 E ) T^2 + E}
\label{eq:el_equals_pl_5}
\end{equation}

The equivalent elastic modulus $M_T$ introduced in Eq. \eqref{eq:el_equals_pl_4} is the harmonic mean of $K/T^2$ and $3\mu$ and represents the coefficient of proportionality between the drop in von Mises stress and the increase in equivalent plastic strain during a {\color{black}plastic burst}. It can only be used when the system has no increase in its total strain. In uniaxial tension ($T$=1/3),  $M_T$ becomes $E$, as stated in Eq. \eqref{eq:ds}. In pure shear ($T$=0), $M_T$ becomes $3\mu$. When $T$ tends to infinity (pure hydrostatic state), $M_T$ becomes equivalent to $K/T^2$. Combining Eq. \eqref{eq:SvM_incrementHomogeneous} and \eqref{eq:el_equals_pl_5}, the relationship between $\discont p$ and $\Delta p_\text{min}$ is found:

\begin{equation}
\discont p = \frac{3 \mu + H}{M_T+H} \Delta p_\text{min}
\label{eq:dpTriaxiality}
\end{equation}

Finally, in uniaxial tension, Eq. \eqref{eq:dpTriaxiality} reduces to Eq. \eqref{eq:dp}, and in pure shear, $\discont p$ is equal to $\Delta p_\text{min}$, as found in the simulations. The model response is thus entirely understood under homogeneous deformation, for any triaxiality.

\subsection{Tensile test on the dogbone geometry}\label{sec:HeteroSection}

\begin{figure}[!h]
\centering
\includegraphics[height=0.6\textheight]{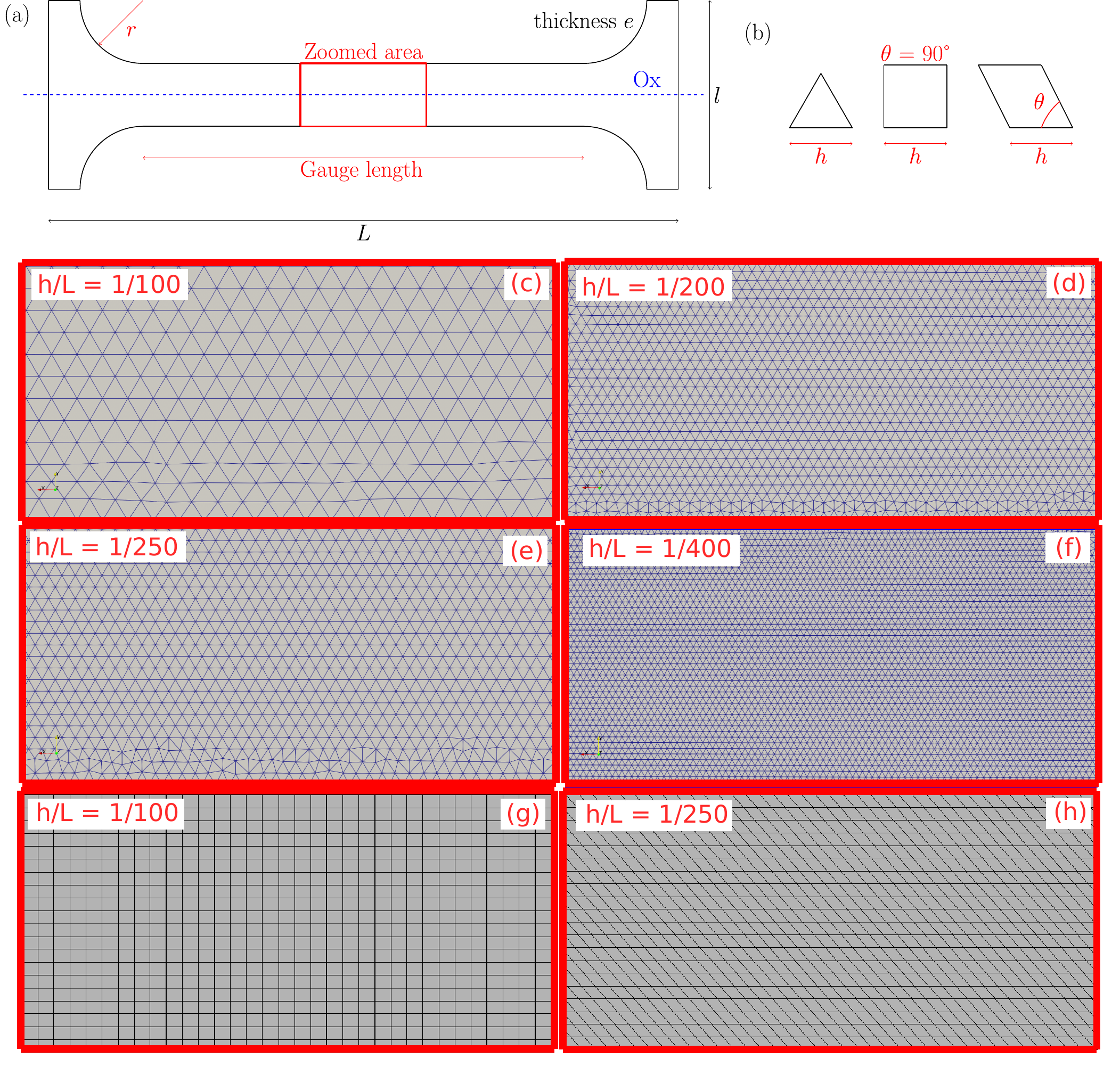}
\caption{(a) Dogbone geometry used in the reference simulation. Dimensions used for the reference simulation are: $L$~=~20, gauge length~=~14, $l$~=~6, $e$~=~0.25, $r$~=~2. The element size is denoted $h$. (b) Meshes used for testing the model (tetrahedral, hexahedral and inclined hexahedral). Representation of the FE meshing in the zoomed area of (a) for (c) a tetrahedral meshing with $h/L$ = 1/100, (d) a tetrahedral meshing with $h/L$ = 1/200, (e) a tetrahedral meshing with $h/L$ = 1/250, (f) a tetrahedral meshing with $h/L$ = 1/400, (g) a hexahedral meshing with $h/L$ = 1/100, (h) an inclined hexahedral meshing with $h/L$ = 1/250. The mesh used in the tetrahedral cases discussed after have a $h/L$ ratio of 1/400.}
\label{fig:SampleDogboneMesh}
\end{figure}

By applying the model to more complex geometries, the macroscopic response becomes very different from the homogeneous case. The example of a 3D dogbone flat specimen is described in this section, whose precise geometry and dimensions are depicted with several 3D meshes tested in Fig. \ref{fig:SampleDogboneMesh}. Simulations are performed in 3D and there are 12 elements in the thickness for the finest mesh. The actual mesh used in this section consists of linear tetrahedra and is depicted in Fig. \ref{fig:SampleDogboneMesh}~(d). The impact of using the other meshes will be discussed in the following section. Uniaxial displacement is imposed on the left and right faces, which are considered clamped (no transverse displacements), and the other faces are free of forces. The elastic and plastic properties of the material, as well as the boundary conditions and the value of the plastic threshold $\Delta p_\text{min}$, are the same as those presented in section \ref{sec:1DTensile}.

\begin{figure}[!h]
\centering
\includegraphics[height=0.85\textheight]{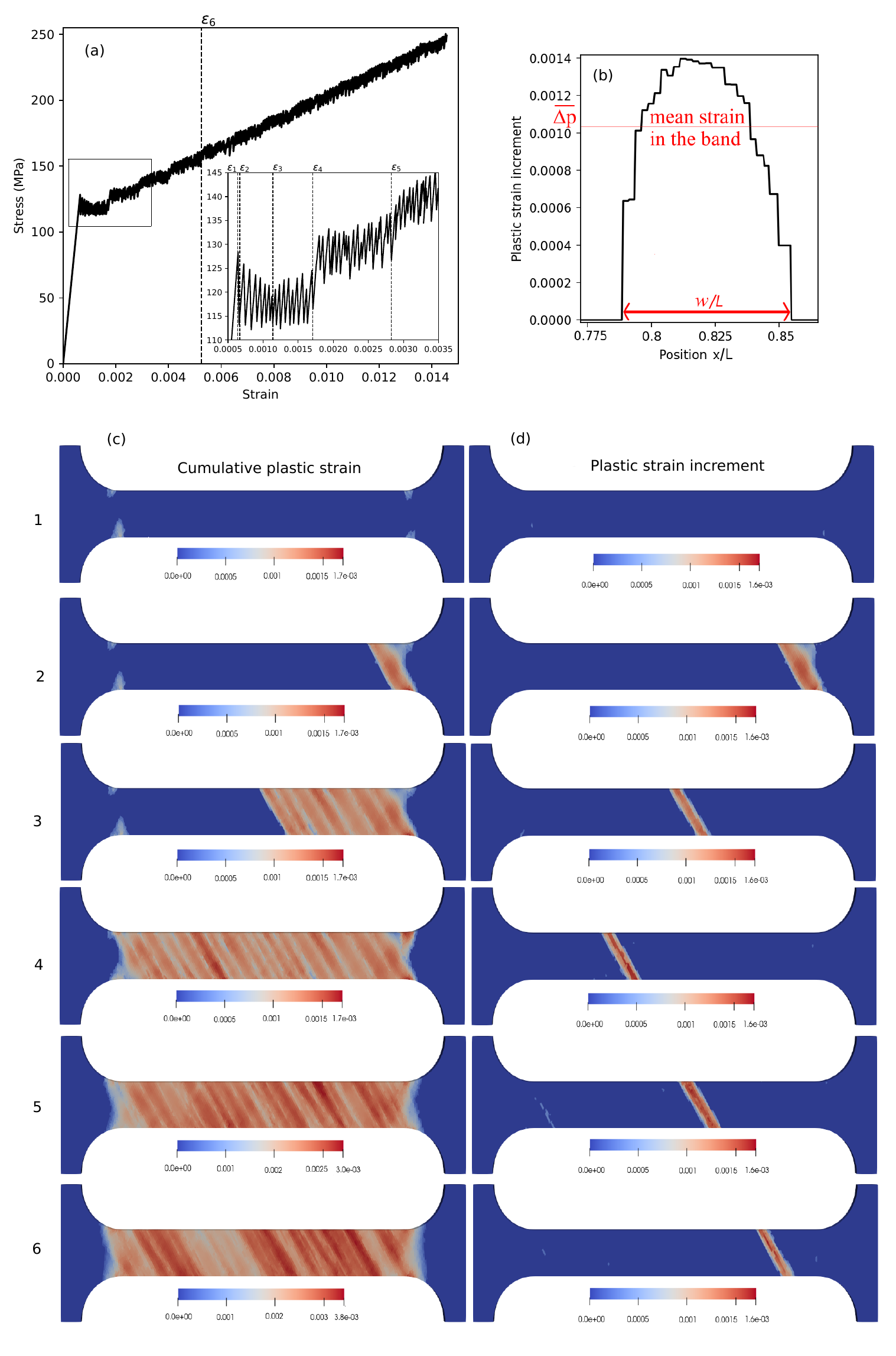}
\caption{(a) Evolution of axial stress as a function of axial strain, obtained by integration over the gauge length of the specimen. (b) Evolution of the plastic strain increment alongside the $O_x$ axis when a plastic {\color{black}burst} occurs at a given numerical step. $w$ is the band width, $L$ is the specimen length and $\overline{\Delta p}$ is the mean plastic strain in the band. (c) Series of cumulative plastic strain map and (d) series of plastic strain increment over the dogbone specimen for 6 given strain levels indicated in (a).}
\label{fig:InhomogeneousResponse}
\end{figure}

The tensile curve is drawn in Fig. \ref{fig:InhomogeneousResponse} (a), obtained by volume averaging of $\varepsilon_{xx}$ and $\sigma_\text{xx}$ over the gauge length, accompanied by maps of cumulative plastic deformation (c) and the corresponding increment (d) at several mean strain levels. Unlike the 1D case, deformation is not homogeneous along the specimen. Plastic strain bands of finite width appear. The width of these bands, defined as the length of elements along the $O_x$ axis that yield at a given increment, is much larger than the width of an element $h$ (see Fig. \ref{fig:InhomogeneousResponse} (b)). The mean plastic strain increment in the deformation band is an order of magnitude greater than the plastic threshold $\Delta p_\text{min}$.

On the stress-strain curve in Fig. \ref{fig:InhomogeneousResponse} (a), it can be observed that, after an elastic regime, serrations are present in the curve. Each serration corresponds to the appearance of a macroscopic plastic strain band on the specimen. From $\varepsilon_1$~=~0.06\% to $\varepsilon_4$~=0.17\% strain, a stress plateau is observed, similar to materials exhibiting static strain aging inducing a Lüders plateau. Beyond 0.17\% strain, the serrations continue. The plateau effect tends to disappear as the strain increases, and the average stress evolution tends to align with the linear hardening rate $H$. This second behavior is analogous to a type B or C PLC effect~\citep{Yilmaz2011,Lamari2024}, notably observed in materials exhibiting dynamic strain aging. Such a behavior is also similar to a well known stick-slip behavior \citep{Perfilyev2013}, as observed in slip or velocity-weakening friction laws \citep{Geus2022}.

The features of the stress-strain curve can be understood by analyzing the dynamics of appearance of plastic bands. An animation of the evolution of cumulative plastic strain map during the simulation is available \href{https://www.youtube.com/watch?v=nW0KEG-W4jk}{here}~\citep{Kerfriden_2023}. More videos are available in the following \href{https://zenodo.org/records/14266823}{zenodo repository} (10.5281/zenodo.14266823). Snapshots of cumulative plastic strain and plastic strain increment maps are respectively given in Fig. \ref{fig:InhomogeneousResponse} (c) and (d). The first band nucleates at one fillet of the specimen due to a local von Mises stress concentration. Just before the first plastic event, a difference of 20~MPa exists between these zones and the central part of the gauge length. These elements are thus the first to exceed the upper yield surface and to deform plastically, as shown at $\varepsilon_1$ in Fig. \ref{fig:InhomogeneousResponse} (c) and (d). At higher deformation $\varepsilon_2$~=~0.07\%, they induce the instantaneous formation of the first traversing band. This first band induces the yield point phenomenon visible on the tensile curve. A cut of this first band alongside the $O_x$ axis is shown in Fig. \ref{fig:InhomogeneousResponse} (b). For each band, we define the mean plastic strain, denoted $\overline{\Delta p}$, and the width of the plastic band, denoted $w$, corresponding to the length of elements along the $O_x$ axis that yield. More precisely, if $L_w$ is the starting position of the band on the specimen length, $l$ the width of the specimen and $e$ its thickness $\overline{\Delta p}$ is calculated by:
\begin{equation}
\overline{\Delta p} = \frac{1}{w} \int_{L_w}^{L_w + w} \discont p \left(x,\frac{l}{2},\frac{e}{2}\right) dx
\label{eq:pmean}
\end{equation}

It is noteworthy that $\overline{\Delta p}$ is always higher than $\Delta p_\text{min}$ (here 5 times higher for the first band). It is recalled that only plastic strain increment higher than $\Delta p_\text{min}$  are accepted in the proposed model. At the edges of the first band, there is a local stress concentration, causing the second band to nucleate next to it. This correlated nucleation of bands persists, as illustrated at $\varepsilon_3$~=~0.11\%. The phenomenon continues until the entire specimen plastically deforms for the first time, marking the end of the initial plateau on the tensile curve. This step corresponds to $\varepsilon_4$~=~0.17\%. Then, in contrast to the Lüders phenomenon, bands still appear along the gauge length until the end of the simulations, as illustrated for $\varepsilon_5$~=~0.28\% and $\varepsilon_6$~=~0.52\%. The localization of consecutive bands becomes more random, and they do not occur at constant stress in the presence of hardening.

The convergence of the model is good compared to the continuum model. The reference simulation has been done with time-discontinuous plasticity and with continuum plasticity (i.e., with $\Delta p_\text{min}$ = 0). The computation time for the former is five times that of the latter. Additionally, the Newton solver used in the overall problem resolution converges quadratically once the set of Gauss points that will yield and those that will remain in the elastic regime are determined.

{\color{black} Simulations can be run under force-controlled conditions by replacing displacement with traction boundary conditions. In this case, plastic deformation occurs in bursts at constant force, producing successive plateaus in the tensile curve, similar to force-controlled tests on alloys with the PLC effect~\cite{Guillermin2023} and as observed in micropillar compression~\cite{Uchic2009}. Each plateau on the stress curve corresponds to plastic deformation bursts occurring within a single numerical step across the entire gauge length.}

\subsection{Statistical analysis methodology}\label{sec:Methodology}

\begin{figure}[!h]
\centering
\includegraphics[width=0.8\textwidth]{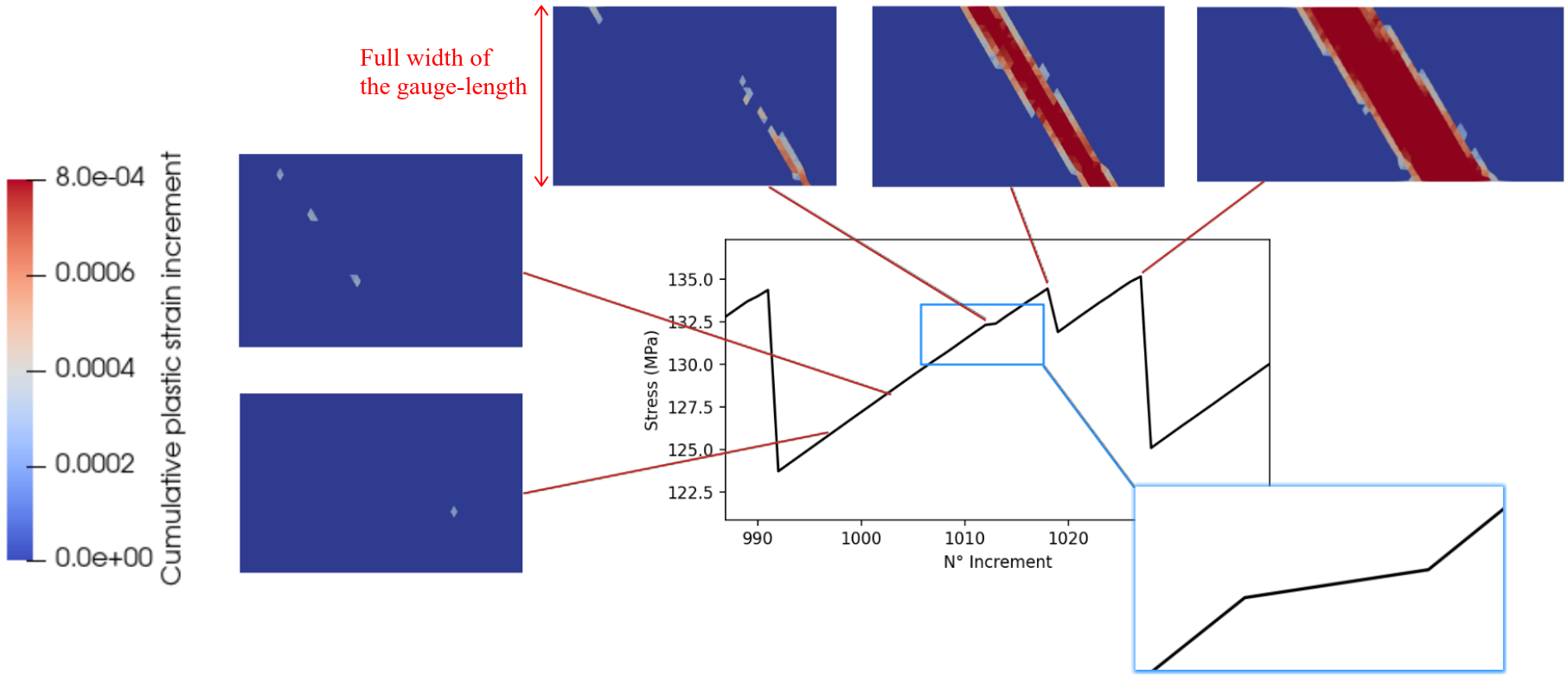}
\caption{Map of plastic strain increment in a region of the specimen for five given plastic events. The instant of their occurence is indicated on the zoomed part of the tensile curve from Fig. \ref{fig:InhomogeneousResponse}. The map height corresponds to the full width of the gauge length.} 
\label{fig:NomenclaturePlasticEvents}
\end{figure}

A close-up view of the tensile curve from Fig. \ref{fig:InhomogeneousResponse} is depicted in Fig. \ref{fig:NomenclaturePlasticEvents}, alongside maps of plastic strain increment across the specimen corresponding to the indicated time. Plastic events can occur in two forms. They can either form bands extending over the whole cross-section, as discussed in the previous section, or one/few elements can deform individually. The first kind are called plastic bands, while the second will be referred to as small plastic events hereafter. Their respective impact on the tensile curve is different. While the plastic bands create a drop in stress, the small plastic events only leads to a change in slope, often barely perceptible to the naked eye.

To capture the effect of each plastic events on the stress curves, we calculate at each increment $n$ of the simulation the quantity $\Delta \sigma = -\left( \sigma_\text{xx}^{n+1} - \sigma_\text{xx}^{n}\right)+ E \Delta\varepsilon_{xx}$, with $\Delta\varepsilon_{xx} = \varepsilon_{xx}^{n+1} - \varepsilon_{xx}^n$. Stress and strain values along $O_x$ are obtained through volumic integration over the gauge length. During the simulation, $\Delta\varepsilon_{xx}$ (obtained by integration over the volume of the gauge length) remains almost constant at $3 \cdot 10^{-6}$. The quantity $\Delta \sigma$ is zero only if there is no plastic event in the gauge length. Otherwise, it is positive. Thus, $\Delta \sigma$ is an adequate measure of the effect of plastic deformation on the average stress. 

\begin{figure}[!h]
\centering
\includegraphics[width=1.0\textwidth]{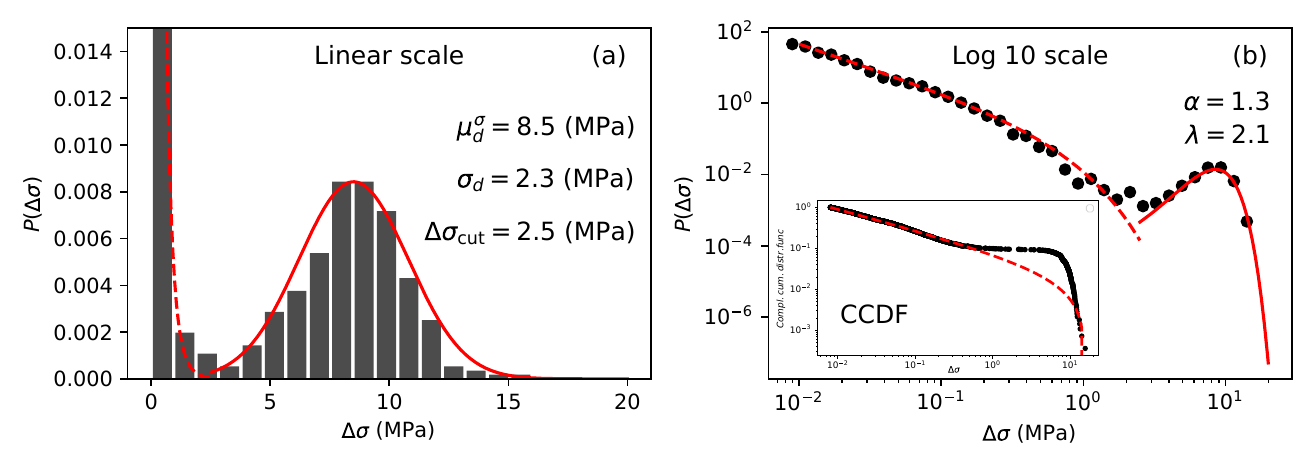}
\caption{Statistics of stress drops (black) determined from the reference simulation in (a) linear scale and (b) logarithmic scale. In red is indicated the modeled distribution using Eq. \eqref{DistributionExpression} (the dashed line is the truncated power law, while the solid line is the Gaussian distribution). The inset of (b) shows the complementary cumulative density function (CCDF) and the power law. There are 5922 events in the data set.}
\label{fig:StatisticsDoubleDistribution}
\end{figure}

The distribution of $\Delta \sigma$ obtained in the simulation detailed in section \ref{sec:HeteroSection} is presented in Fig. \ref{fig:StatisticsDoubleDistribution} in black in both linear (a) and logarithmic (b) scales. The inset of (b) shows the complementary cumulative density function (CCDF) in logarithmic scale. The distribution of $\Delta \sigma$ consists of two distinct parts: above a stress drop of $\Delta \sigma \approx$~2.5~MPa, it is bell-shaped, as observed on the linear scale; and below 2.5~MPa, it has a power law shape, visible on the logarithmic scale by its linear appearance. On the logarithmic scale, the shape of the distribution is reminiscent of supercriticality \citep{ZhangSalman2020,Richeton2005}, where the distribution of events follows a power law, except for extreme events that are overrepresented due to a favoring mechanism (Dragon Kings phenomenon \citep{Sornette2012}). However, the distribution of big events following a well-defined bell curve pleads for a difference in the physical origin of the two populations of plastic events. Therefore, in this paper we will characterize both parts of the distributions independently. The small events will be analyzed using a truncated power law. The big events, having a well-defined bell shape, will be analyzed using a Gaussian distribution. The overall distribution of $\Delta \sigma$ is divided into two parts using a cut-off value, denoted $\Delta \sigma_\text{cut}$, determined empirically from the simulation (here 2.5~MPa). The probability density function (PDF) of the stress drop $P \left( \Delta \sigma \right)$ is given by:
\begin{flalign}
& \text{if } \Delta \sigma > \Delta \sigma_\text{cut}: \quad P \left( \Delta \sigma \right) = H_d \cdot \left(\frac{1}{\sigma_d \sqrt{2 \pi}}\right) \cdot \exp\left(-\frac{(\Delta \sigma - \mu_d^{\sigma})^2}{2 \sigma_d^2}\right) \\ 
& \text{if } \Delta \sigma < \Delta\sigma_\text{cut}: \quad P \left( \Delta \sigma \right) =  C_d \cdot \Delta \sigma^{-\alpha} e^{-\lambda \Delta \sigma}
\label{DistributionExpression}
\end{flalign}
In Eq. \eqref{DistributionExpression}, the parameters $\mu_d^{\sigma}$ and $\sigma_d$ are the mean and standard deviation of the Gaussian distribution, the parameter $\alpha$ is the exponent of the power law, $\lambda$ is a parameter that accounts for the deviation from pure power law due to the non-infinite system size~\citep{ZhangSalman2020}, and $H_d$ and $C_d$ are two constants quantifying the respective weight of the two contributions to the PDF. To obtain those parameters, the distribution of small events (< $\Delta \sigma_\text{cut}$) is first calculated. The exponent $\alpha$ and parameter $\lambda$ are then extracted using the maximum likelihood method~\citep{ZhangSalman2020,Aaron2009}, which was implemented using the "powerlaw" library in python \citep{Alstott2014}. The latter library also allows to evaluate the likelihood of two given distributions to represent a distribution. Truncated power law was better suited than power law, exponential, stretched exponential and log-normal distribution to model the distribution. To obtain $\mu_d^{\sigma}$ and $\sigma_d$, the distribution of big events(> $\Delta \sigma_\text{cut}$) is calculated. The method of least squares was then used, with a value of $H_d$ = 1 being imposed. Finally, the values of $H_d$ and $C_d$ are chosen so that the integral of the two modeled distributions is equal to the integral of the corresponding part of the distributions of all events.

A problem with the choice of the Gaussian function to model the PDF of $\overline{\Delta p}$ and $w$ is the nonzero probability of negative events, which is unphysical. However, this probability is negligible. Furthermore, simple functions with strictly positive support, such as log-normal distributions, are unsuitable to model the observed distribution because they are long-tailed and non-symmetric. Gaussian distribution was therefore used.

For simplicity, we will focus our analysis on the $\mu_d^{\sigma}$ and $\alpha$ parameters to characterize the evolution of each portion of $P \left( \Delta \sigma \right)$ as we change the model parameters. Respectively, $\mu_d^{\sigma}$ represents the mean stress drop associated with plastic bands, and $\alpha$ characterizes the nature of the plastic flow associated with small plastic events. Thus, they are the only parameters of $P \left( \Delta \sigma \right)$ that will be analyzed afterwards.

\begin{figure}[!h]
\centering
\includegraphics[width=1.0\textwidth]{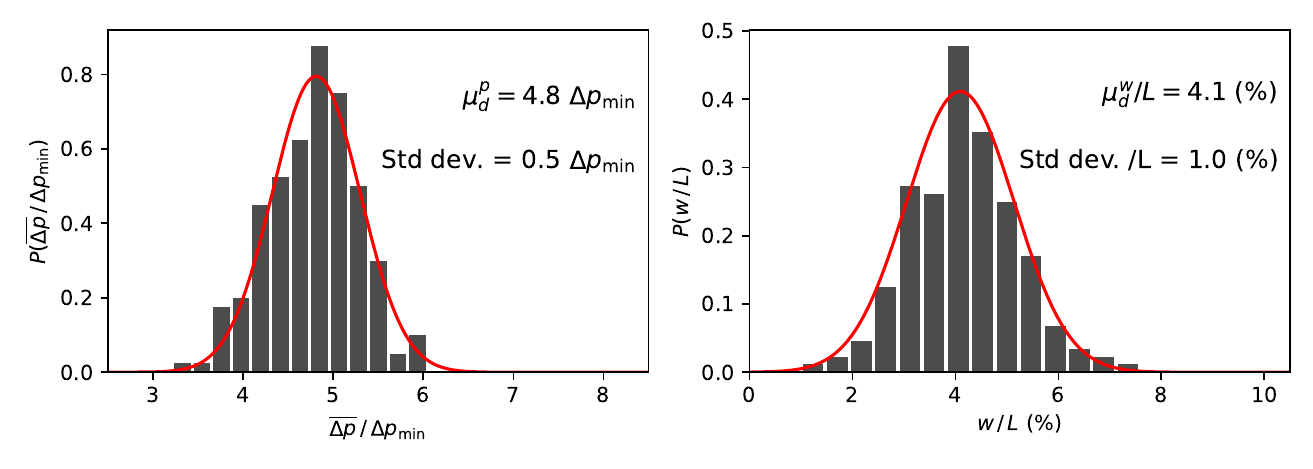}
\caption{Statistics of (a) mean plastic strain and (b) band widths (black bar chart) determined from the reference simulation. In solid line is indicated the modeled distribution using a Gaussian distribution. There are 187 events in the data set.}
\label{fig:PlasticStatistics}
\end{figure}

Finally, the plastic activity over the $O_x$ axis has also been analyzed. The distributions of mean plastic strain $\overline{\Delta p}$ (as defined in Eq. \eqref{eq:pmean}) and the width of the plastic bands $w$ (see Fig. \ref{fig:InhomogeneousResponse} (b) for a graphical representation) have been determined during the simulations. With this methodology, all plastic bands are characterized, since they all cross the $O_x$ axis, but only a fraction of the small plastic events are captured, i.e. the ones situated on the $O_x$ axis. To further remove from the statistics the effect of small plastic events, a threshold is done on the value of $\overline{\Delta p}$, keeping only those above the threshold. Typically, a plastic event is considered a band when $\overline{\Delta p} > 3 \Delta p_\text{min}$. Hence, only the plastic activity of plastic bands are analyzed afterwards. The distributions of $\overline{\Delta p}$ and $w$ obtained from the reference simulation have a Gaussian shape, as shown in Fig. \ref{fig:PlasticStatistics} for the reference simulation. This is consistent with the distribution of $\Delta \sigma$ which had a Gaussian part due to plastic bands. Therefore, distributions of $\overline{\Delta p}$ and $w$ are also modeled with Gaussian distribution, as described in Eq. \eqref{DistributionExpression}. The mean values of $\overline{\Delta p}$ and $w$ obtained by this method are respectively denoted  $\mu_d^{p}$ and $\mu_d^{w}$.

\section{Assessment of the model} 

In this section, we focus on the behavior of plastic bands. The effects of smaller-scale plastic events on elastic properties are analyzed in the subsequent section. If not explicitly stated otherwise, the parameters used in the simulations will be the same as those used in the preceding sections and are provided in Table \ref{tab:params}. Complete datasets discussed after can be found at \href{https://zenodo.org/records/14266823}{10.5281/zenodo.14266823}.

\begin{table}[!h]
\centering
\begin{tabular}{|l|l|}
\hline
\textbf{Parameter} & \textbf{Value} \\
\hline
Initial yield stress $\sigma_{\text{ys}}$ & 100 MPa \\
Linear hardening rate $H$ & 10 GPa \\
Young's modulus $E$ & 200 GPa \\
Poisson's ratio $\nu$ & 0.3 \\
Plastic threshold $\Delta p_\text{min}$ & $2 \times 10^{-4}$ \\
Prescribed deformation increment $\Delta \varepsilon_{xx} = \Delta u /L$ & $3 \times 10^{-6}$ \\
Ratio mesh size over specimen length $h/L$ & $0.5\%$ \\
\hline
\end{tabular}
\caption{Reference material and simulation parameters.}
\label{tab:params}
\end{table}

\subsection{Mesh sensitivity}

The effect of meshing, with some examples displayed in Fig. \ref{fig:SampleDogboneMesh}, is studied in this section. First, the effect of mesh size is tested on linear tetrahedral elements. Then, the effect of mesh type (tetrahedron vs hexahedron), and the angles of hexahedra with respect to the tensile direction are analyzed. 

\begin{figure}[!htb]
\centering
\includegraphics[width=1\textwidth]{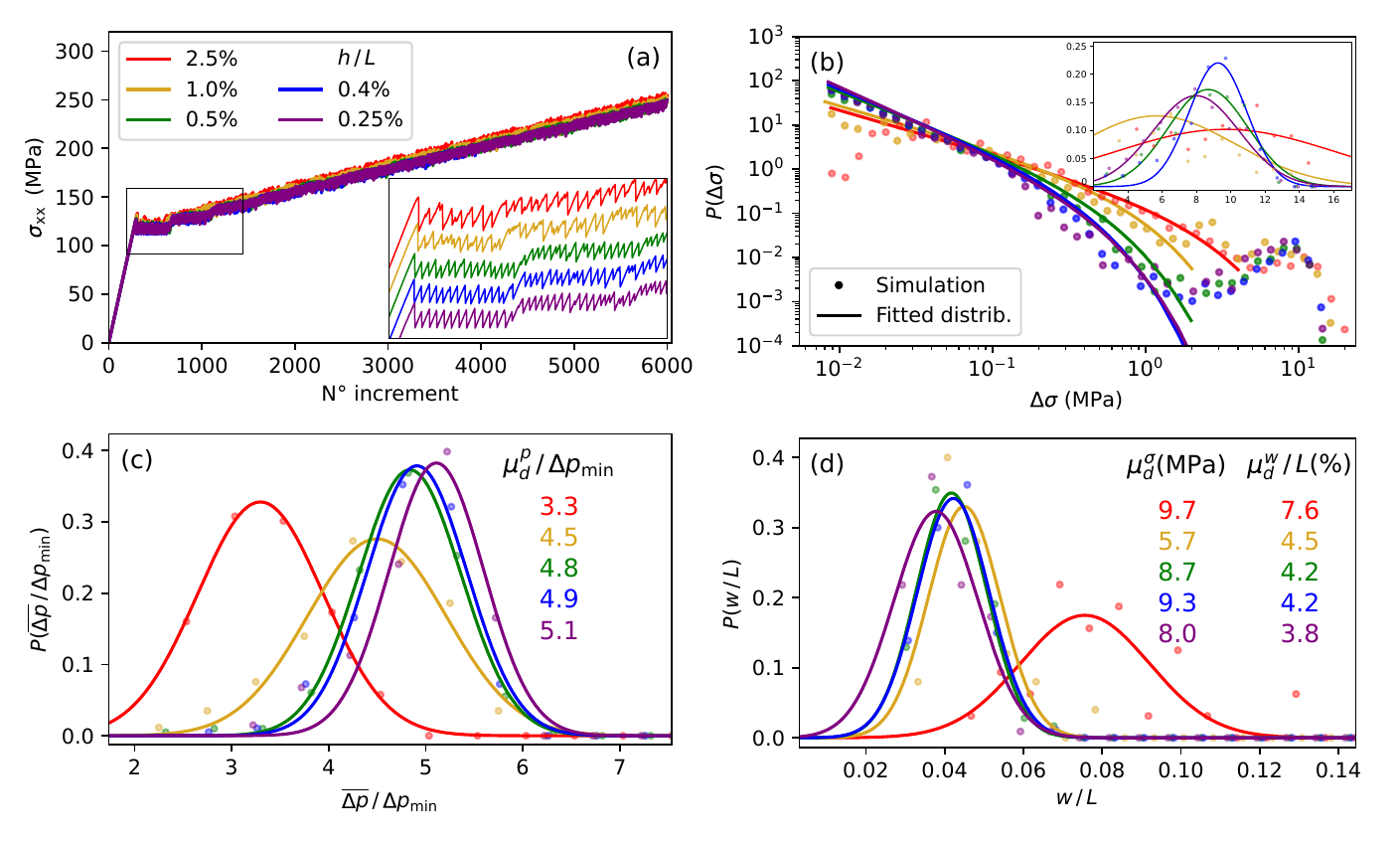}
\caption{Effect of varying the ratio of mesh size over specimen length $h/L$ on (a) the tensile curves, (b) the statistics of stress drops, (c) the statistics of mean plastic strain and (d) the statistics of band widths. In (b), (c) and (d), the dots are representative of the probability density function determined from the simulation and the solid line is the modeled distribution. The color code is indicated in (a). In (a), the tensile curves are shifted in the inset for visibility purposes. In (b), the distributions are indicated in linear scale in the inset.}
\label{fig:TailleMaille}
\end{figure}

Various ratios of mesh size $h$ over specimen length $L$ are tested. Specimen gauge length and heads have the same element size. The same conditions are used as described in section \ref{sec:HeteroSection}. The tensile curves are given in Fig. \ref{fig:TailleMaille} (a). The effective yield strength for $h/L$ = 2.5\% is 134 MPa. It is superior to the initial yield strength $\sigma_{\text{ys}}$ because in time-discontinuous plasticity the yield surface does not define the stress at which plasticity begins (see Fig. \ref{fig:ThickYieldSurface}). The effective yield strength decreases to 132 MPa for $h/L$ = 1.5\%, and finally converges at 127 MPa for meshes finer than 0.5\%. The average flow stress during the first plateau is 117 MPa and is mesh independent. The serrations are more varied in intensities for coarser meshes, while they become more periodic for meshes finer than 0.5\%.

The PDF of stress drops $\Delta \sigma$ extracted from the simulations are given in Fig. \ref{fig:TailleMaille} (b) in logarithmic and linear scales. The mean value of stress drops $\mu_d^\sigma$ linked with plastic bands are indicated in Fig. \ref{fig:TailleMaille} (d). The modeled distribution are also shown. Looking at the statistics of stress drops reveals that the distribution of $\Delta \sigma$ converges for a mesh size below 1/100 of the specimen size, with the difference in the average stress drop $\mu_d^\sigma$ being about 1~MPa, which is smaller than the standard deviation of the distribution ($\approx$~2.6~MPa for purple distribution).

In Fig. \ref{fig:TailleMaille} (c) the distributions of mean plastic strain $\overline{\Delta p}$ of bands obtained in the five simulations are indicated, alongside the average values of these distributions. All values are normalized by $\Delta p_\text{min}$. The band strain tends to be smaller for larger mesh elements. The average value $\mu_d^p$ converges as the mesh size decreases, approaching 5 times $\Delta p_\text{min}$. In Fig. \ref{fig:TailleMaille} (d) the distributions of band width $w$ obtained in the five simulations are indicated, alongside the average values of these distributions. All values are normalized by the length of the specimen length $L$. The band width is larger for coarser meshes. The average value $\mu_d^w$ converges as the mesh size decreases, approaching 4\% of the specimen length L (the standard deviation being 1\% of L in comparison). Overall, this analysis demonstrates that all relevant distributions (not just the average values) converge for mesh sizes below 0.5\% of the total specimen length.

\begin{figure}[!htb]
\centering
\includegraphics[width=1\textwidth]{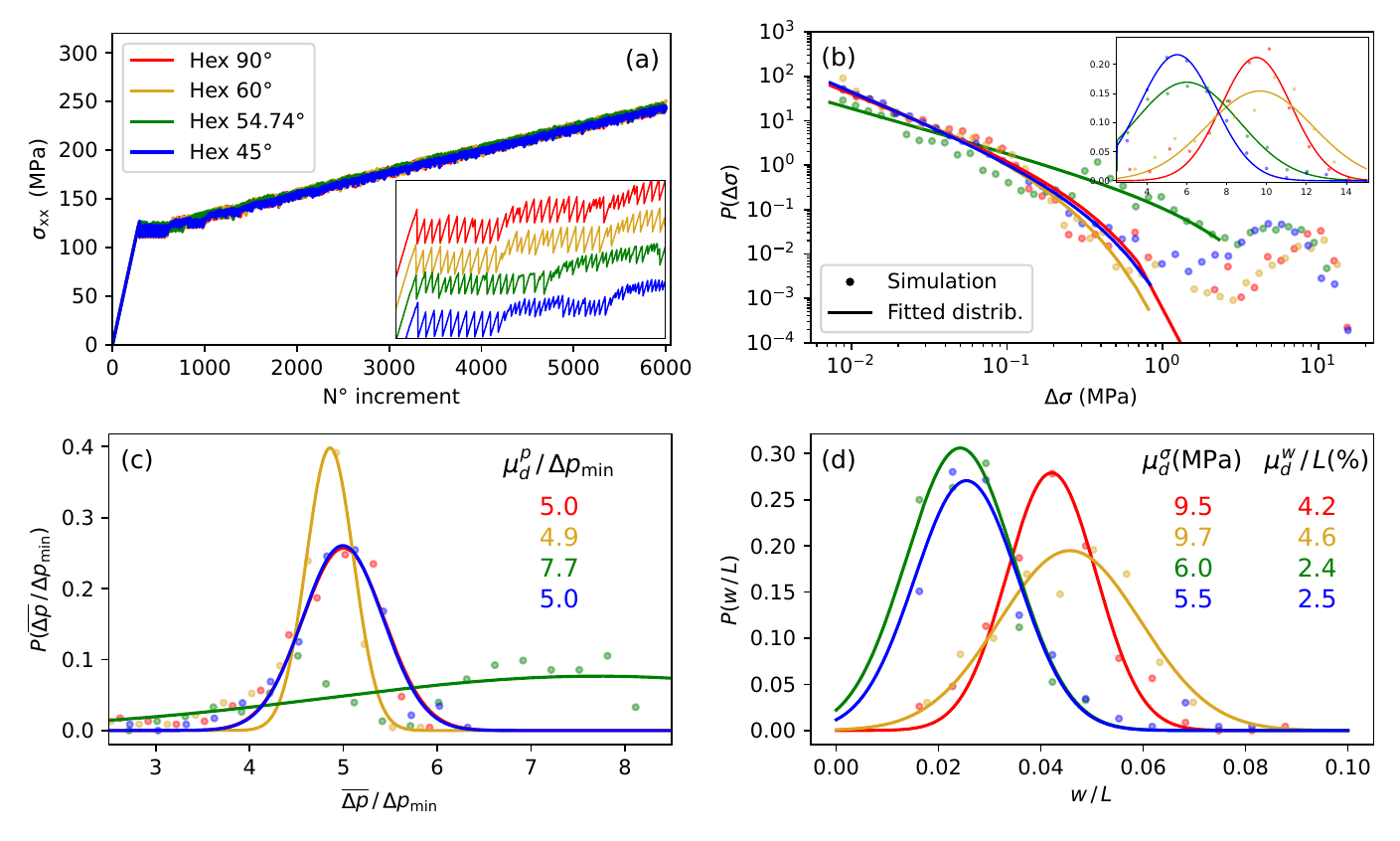}
\caption{Effect of varying the orientation $\theta$ of hexahedral meshes on (a) the tensile curves, (b) the statistics of stress drops, (c) the statistics of mean plastic strain and (d) the statistics of band widths. In (b), (c) and (d), the dots are representative of the probability density function determined from the simulation and the solid line is the modeled distribution. The color code is indicated in (a). In (a), the tensile curves are shifted in the inset for visibility purposes. In (b), the distributions are indicated in linear scale in the inset.}
\label{fig:Hexahedre}
\end{figure}

The results of the simulations with hexahedral elements are given in Fig. \ref{fig:Hexahedre} with the same format as in Fig. \ref{fig:TailleMaille}. The ratio $h/L$ is set to 1/400 for hexahedral meshes, with $h$ defined in Fig. \ref{fig:SampleDogboneMesh}. Four orientations $\theta$ of hexahedra have been tested, with angles with respect to the tensile axis ranging from 45° to 90° ($\theta$ indicated in Fig. \ref{fig:SampleDogboneMesh}). The angle 54.74° has been selected since it is the orientation observed in isotropic elastoplastic materials of plastic strain localization bands \citep{Maziere2015}. The tensile curves of the 4 simulations (Fig. \ref{fig:Hexahedre} (a)) are in good qualitative agreement. Looking at the serrations, they appear more regular when the mesh is oriented at 90° or 60° than for smaller angles. 

The stress drop distributions (Fig. \ref{fig:Hexahedre} (b)) reveals the same division, with 90° and 60° Gaussian distribution centered on 9.6~MPa and 54.74° and 45° distributions centered on 5.7~MPa. The same division can be observed in the distributions of band width in Fig. \ref{fig:Hexahedre} (d), with overall smaller bands for 54.74° and 45° orientations. However, the mean plastic strain distributions reveals that while 45° mesh phenomenology is very similar to 60° and 90° meshes (Gaussian distribution centered around $\mu_d^p$~=~5$\Delta p_\text{min}$), the 54.74° distribution is significantly different. In this case, $\overline{\Delta p}$ distribution (green dots in Fig. \ref{fig:Hexahedre} (c)) contains two Gaussian-like parts, one centered on 4.5$\Delta p_\text{min}$ and the other on 7.7$\Delta p_\text{min}$. The mean of the 54.74° distribution (calculated on simulation data, not on the Gaussian model) is 6.0$\Delta p_\text{min}$, which is significantly higher than other distributions (standard deviation $\approx$ 0.6$\Delta p_\text{min}$).

The effect of mesh orientation on overall specimen response can be understood similarly to the effect of mesh orientation on the simulations of the Lüders phenomenon studied by \citep{Maziere2015}, within the framework of a classical continuum plasticity model and using a softening material. In this work, it is shown that meshes oriented at 54.74° allowed for discontinuities in the strain tensor (i.e. infinitesimal localization bands), which were not observed for meshes oriented at 90° (where band fronts were more diffuse). This is a well-known feature of Lagrange linear and quadratic interpolation functions which allow for strain discontinuities along edges but not inside the elements. Spurious effects arise when elements are crossed by discontinuity lines such as strain bands. This could be improved using Galerkin discontinuous types of elements. In our time-discontinuous model, similar features are observed, although bands with a width of one mesh element never appear while they were observed in \citep{Maziere2015} for Lüders modeling. A mesh that permits perfect plastic localization alters the statistics of plastic strain, stress drops and band widths. Meshes with a 45° orientation also seem to facilitate the accommodation of more localized bands, but to a lesser extent. Fortunately, when the mesh is more randomly oriented with respect to the ideal localization band orientation, the statistics of the bands seem to converge, and the convergence values (mean and standard deviation) match those found for tetrahedral meshes. Therefore, for the model to yield a mesh-independent response, meshes must be sufficiently refined and not oriented in a way that favors the formation of localized bands.

\subsection{Effect of the strain increment $\Delta \varepsilon_{xx}$}

In the simulation, axial displacement is imposed at the end of the specimens. This displacement can be linked with a increment of axial strain $\Delta \varepsilon_{xx}$ in the gauge length, which is constant during the whole simulation. In this section, we examine the convergence of the model with respect to $\Delta \varepsilon_{xx}$. The mesh consists of tetrahedra with a $h/L$ ratio of 5\%.

\begin{figure}[!h]
\centering
\includegraphics[width=1\textwidth]{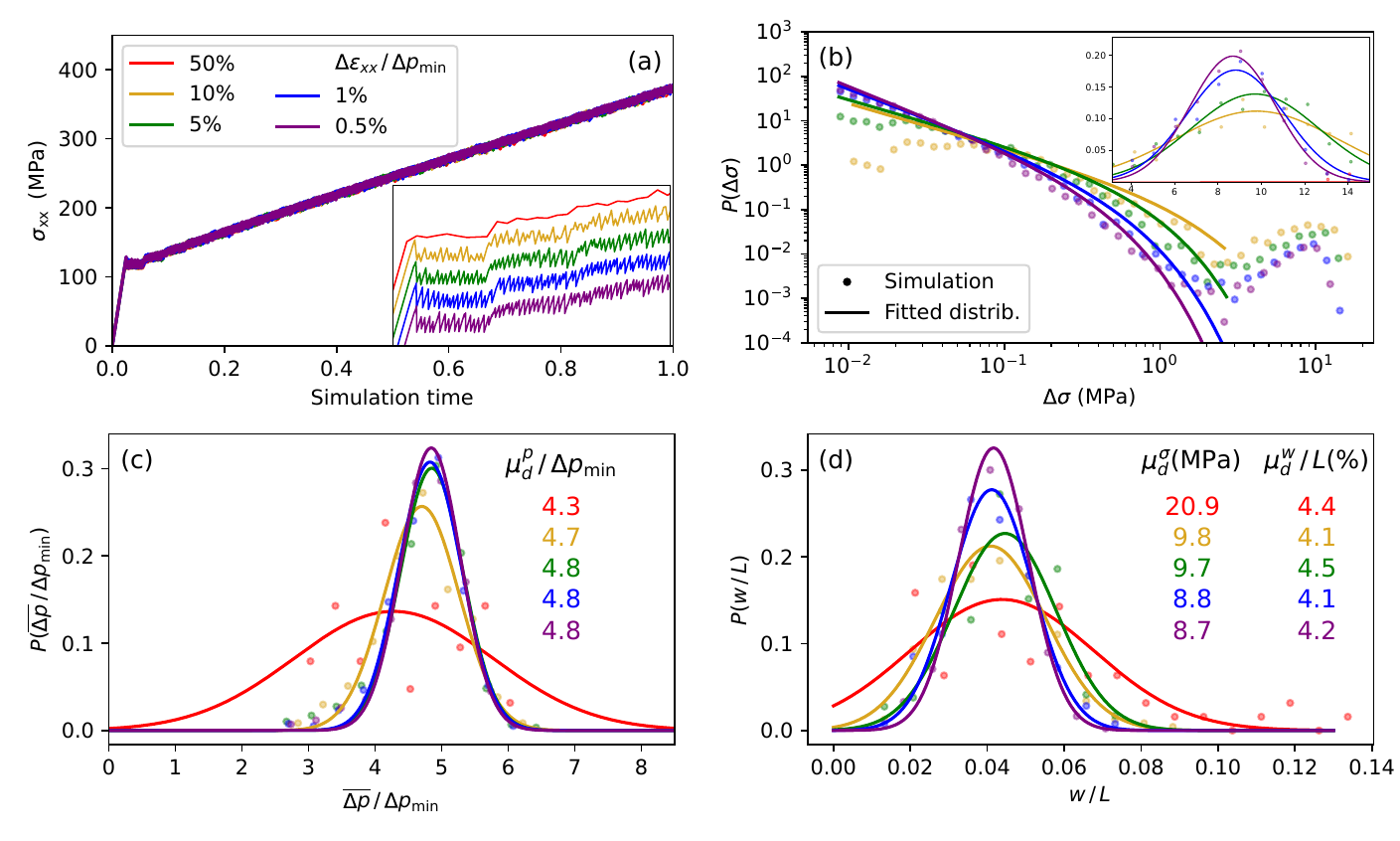}
\caption{Effect of varying the ratio of axial strain increment over plastic threshold $\Delta \varepsilon_{xx}/\Delta p_\text{min}$ on (a) the tensile curves, (b) the statistics of stress drops, (c) the statistics of mean plastic strain and (d) the statistics of band widths. In (b), (c) and (d), the dots are representative of the probability density function determined from the simulation and the solid line is the modeled distribution. The color code is indicated in (a).  In (a), the tensile curves are shifted in the inset for visibility purposes. In (b), the distributions are indicated in linear scale in the inset.}
\label{fig:Epsilon}
\end{figure}

In Fig. \ref{fig:Epsilon} (a), the tensile curves are shown for five simulations. When $\Delta \varepsilon_{xx}$ is too high, no pronounced serration are observed on the tensile curve. As it decreases, clear serrations appear (yellow and green). As $\Delta \varepsilon_{xx}$ decreases further, serrations tend to become more regular. 

The statistics of stress drops, mean plastic strains and band widths are given for the five simulations respectively in Fig. \ref{fig:Epsilon} (b), (c) and (d). For the largest $\Delta \varepsilon_{xx}$, the distribution of stress drops is not displayed as, in the absence of serration on tensile curve, they have little meaning. The distributions of band mean plastic strains and sizes for $\Delta \varepsilon_{xx}$~=~0.5$\Delta p_\text{min}$ reveal that bands appear in this case, despite the absence of serration on the tensile curves, and are very diverse in size and intensity. This apparent contradiction is explained by the fact that while plastic bands appear and tend to decrease the stress of the system, the axial stress increment provided to the system $E \Delta \varepsilon_{xx}$ at each numerical step is high enough to compensate the drop.

When $\Delta \varepsilon_{xx}$ is decreased, the three distributions converge quickly, not only in terms of average values but also in shape. Therefore, the model response converges when the displacement increment imposed at the boundary is small enough. The criterion is for $\Delta \varepsilon_{xx}$ to be below 0.05$\Delta p_\text{min}$. This analysis demonstrates that the presented model converges with respect to the incremental change in boundary conditions.

\subsection{Effect of constitutive parameters} 

\begin{figure}[!h]
\centering
\includegraphics[width=\textwidth]{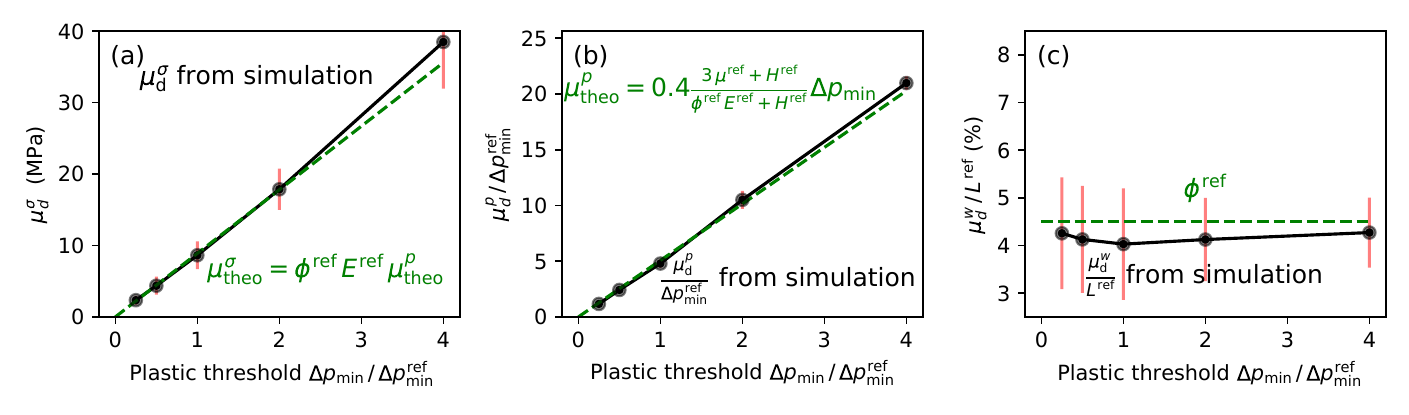}
\vspace{-10pt} 
\includegraphics[width=\textwidth]{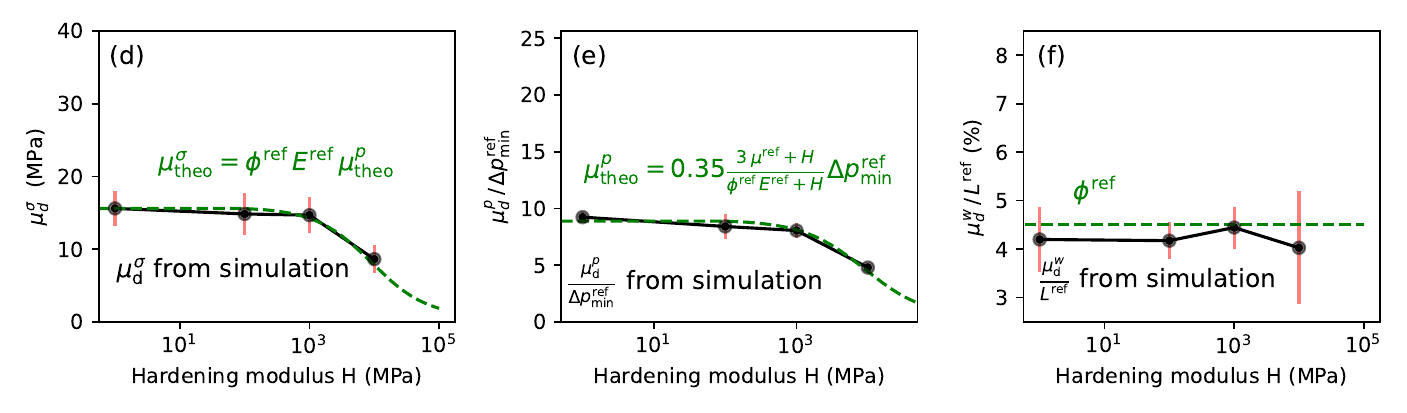}
\vspace{-10pt} 
\includegraphics[width=\textwidth]{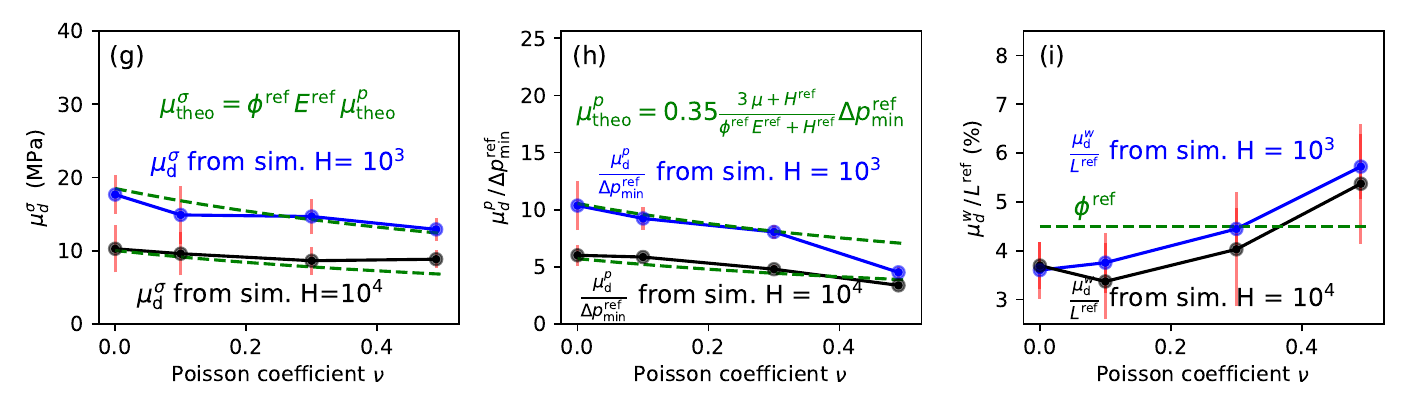}
\vspace{-10pt} 
\includegraphics[width=\textwidth]{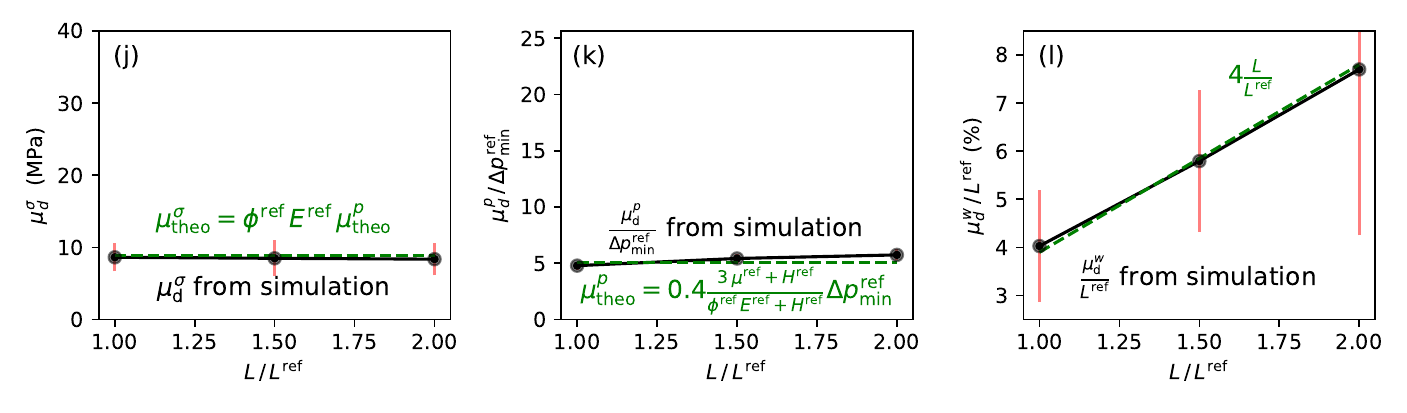}
\vspace{-10pt} 
\caption{Effect of plastic threshold $\Delta p_\text{min}$ on (a) $\mu_d^\sigma$,  (b) $\mu_d^p$ and (c) $\mu_d^w$. Effect of work-hardening  modulus $H$ on (d) $\mu_d^\sigma$,  (e) $\mu_d^p$ and (f) $\mu_d^w$. Effect of Poisson's ratio $\nu$ on (g) $\mu_d^\sigma$,  (h) $\mu_d^p$ and (i) $\mu_d^w$. Effect of gauge length $L$ on (j) $\mu_d^\sigma$,  (k) $\mu_d^p$ and (l) $\mu_d^w$. In black are indicated the results from simulations. For $\nu$, two different values of $H$ were tested. The values found for $H$~=~1000~MPa is indicated in blue, and the values found for $H$~=~10000~MPa is indicated in black. In green dashed line are the theoretical evolutions found using Eq. \eqref{eq:dp_with_phi} and \eqref{eq:ds} using the reference parameters, as written for each figure.}
\label{fig:ModelParametersEffects}
\end{figure}

The effects of constitutive parameters are studied in this section on dogbone specimens with tetrahedral meshes. For conciseness, only mean values of stress drop $\mu_d^\sigma$, band plastic strain $\mu_d^p$ and band size $\mu_d^w$ will be analyzed. Fig. \ref{fig:ModelParametersEffects} (a), (b) and (c) show the effects of $\Delta p_\text{min}$, varying from $5 \times 10^{-5}$ to $8 \times 10^{-4}$ on the evolution of $\mu_d^\sigma$,  $\mu_d^p$ and $\mu_d^w$ determined from simulations (black dots). In this range, both $\mu_d^\sigma$ and $\mu_d^p$ increase linearly with $\Delta p_\text{min}$, while the value of $\mu_d^w$ changes marginally (variation well below the standard deviation of each distribution, indicated by a red lines in Fig. \ref{fig:ModelParametersEffects} (c)).

The linear work-hardening modulus $H$ was varied from 0, to 10000 MPa. The corresponding results for the evolution of $\mu_d^\sigma$,  $\mu_d^p$ and $\mu_d^w$ are given in Fig. \ref{fig:ModelParametersEffects} (d), (e) and (f). For $H$ ranging from 0 to 1000~MPa, no characteristic of the bands evolves. Between 1000 and 10000~MPa, both $\mu_d^\sigma$ and  $\mu_d^p$ decrease by a factor 1.7, while $\mu_d^w$ decreases only by a factor of 1.1. 

Finally, the last model parameter tested is $\nu$. It was varied from 0 to 0.49. Evolution of $\mu_d^\sigma$,  $\mu_d^p$ and $\mu_d^w$ as $\nu$ is varied are given in Fig. \ref{fig:ModelParametersEffects} (g), (h) and (i), for two given values of $H$: 10000~MPa in black, and 1000~MPa in blue. For $H$~=~10000 MPa, varying $\nu$ from 0 to 0.49 decreases $\mu_d^\sigma$ by a factor of 1.2, $\mu_d^p$ by a factor of 1.8 and increases $\mu_d^w$ by a factor of 1.5. For $H$~=~1000 MPa, varying $\nu$ from 0 to 0.49 decreases $\mu_d^\sigma$ by a factor of 1.4, $\mu_d^p$ by a factor of 2.3 and increases $\mu_d^w$ by a factor of 1.5. Therefore, it is clear that when $H$ is larger, the effect of Poisson's ratio on the band plastic strain and stress drop intensity is less significant. It is also apparent that $\nu$ has a significant impact on the band width, and that the effect of $H$, while small, is not insignificant, as a constant factor of 1.1 on $\mu_d^w$ is almost always found when comparing results obtained for $H$~=~10000 MPa and $H$~=~1000 MPa.

While no explanation has been found yet for the dependence of band width with each parameter, it is possible to understand the evolutions of band strain and stress drop. Assuming that the fraction  $\phi$ of the plastic band in the specimen is known and remains constant for each {\color{black}plastic burst}, it is possible to predict the value of the plastic strain in the band: 

\begin{equation}
\discont p = \frac{3 \mu + H}{\phi E+H} \Delta p_\text{min}
\label{eq:dp_with_phi}
\end{equation}

The Eq. \eqref{eq:dp_with_phi}, analogous to Eq. \eqref{eq:dp} and \eqref{eq:dpTriaxiality}, is derived in the same way by replacing $\discont{\tenstwo{\varepsilon}^{\mathrm{p}}}$ by $\phi \discont{\tenstwo{\varepsilon}^{\mathrm{p}}}$ in Eq.~\eqref{eq:NegligibleStrain}. This formula is valid if the stress state in the specimen is tensile and strictly uniform. For a dogbone specimen, the value of $\phi$ for a given band is close to the ratio between the band width and the specimen length $w/L$. As the value of $\phi$ remains a priori unknown in the model, a constant value $\phi^{ref}$~=~4.5\% is chosen for all future calculations, close to the value of $\mu_{d}^w / L$ found in the simulation.

Using the previous Eq. \eqref{eq:dp_with_phi} and \eqref{eq:ds} ($\discont{\sigma_\text{vM}} = - E \discont p$), and assuming a value for $\phi$ = $\phi^{ref}$, it is possible to give an estimate $\mu_{theo}^\sigma$ and $\mu_{theo}^p$ for all simulations (green dashed curves in Fig. \ref{fig:ModelParametersEffects}). A multiplicative factor between 0.35 and 0.4 has been applied to the obtained values to be closer to the simulations results (as indicated in Fig. \ref{fig:ModelParametersEffects} (b), (e), (h) and (k)). This multiplicative factor can be explained by geometric consideration, making the stress state far from uniform in the dogbone specimen. 

Overall, the proportionality factor between $\Delta p_\text{min}$ and both $\mu_{d}^\sigma$ and $\mu_{d}^p$ is well explained, alongside the slightly more complex evolution of those quantities. The small gap between simulation data and modeled evolution can be explained by the non constant value of $\phi$ in simulation, especially as $\nu$ change. All model parameters affect the plastic strain in bands and the consecutive stress drop in a theoretically well-understood manner. Most model parameters do not affect the band width, but some parameters, especially $\nu$, do, for reasons that are not fully understood yet.

\subsection{Effect of gauge length}

To study the impact of geometry on the model response, the gauge length $L$ has been modified from a value $L^{ref}$ (which was the one used before) to a value of 2$L^{ref}$. Fig. \ref{fig:ModelParametersEffects} (j), (k) and (l) display the effects of $L$ on the evolution of $\mu_d^\sigma$,  $\mu_d^p$ and $\mu_d^w$ determined from simulations. The average value of band widths $\mu_d^w$ evolves proportionally to the gauge length, with a factor of proportionality of 4\%, which is the average fraction $\phi$ of the specimen in the plastic band in the reference state (see Fig. \ref{fig:Epsilon} (d)). This means that the ratio $\phi = \mu_d^w/L$ remains constant when $L$ is modified. On the other hand, changing the gauge length has no effect on $\mu_d^\sigma$ and $\mu_d^p$, which was expected from Eq. \eqref{eq:dp_with_phi} and \eqref{eq:ds}, since $\phi$ remains constant as $L$ changes and other parameters do not depend on $L$.

Simulation with dogbone geometries transformed under homothetic dilatation have been performed. Dilatation factors of 2 and 4 have been tested (results not shown here). Again, all dimensions of the bands evolve proportionally with the corresponding dimensions of the specimen (band width with specimen length, band height with specimen width and band thickness with specimen thickness). Other features of the bands remain unaffected by homothetic transformation (band mean plastic strain and stress drop intensity). The model is thus homothetic. This result is consistent with the absence of an internal length in the model. Only a plastic threshold has been introduced, which is not linked with any length.

\section{Discussion}

\subsection{Statistics of stress drop amplitude and frequency}

In the previous sections, it has been demonstrated that the plastic events occur following two distributions: a Gaussian part (big plastic events) and a power law part (small plastic events). The properties of big plastic events alongside the effect of model parameters on them have been extensively analyzed in the previous sections. Here, we analyze the truncated power law distribution. 

In Fig. \ref{fig:TailleMaille}, \ref{fig:Hexahedre} and \ref{fig:Epsilon} (b) the distributions of stress drops in logarithmic scales are represented for the simulations testing the effects of mesh size, orientation, and prescribed displacement at boundaries, respectively. The simulation results (dots) are displayed with the modeled truncated power law (solid line). When the mesh size is too big, or the the axial strain  $\Delta \varepsilon_{xx}$ too large, deviation from power law is observed for stress drop events smaller than $10^{-1}$~MPa. However, when $h/L$ is lower than 0.5\%, and $\Delta \varepsilon_{xx} / \Delta p_\text{min}$ lower than 1 \%, then the stress drop distribution follow indeed a power law for its smaller events. For any orientation of the hexahedral mesh, a good match is found between the stress drop distribution and the power law. 

\begin{figure}[!h]
\centering
\includegraphics[width=\textwidth]{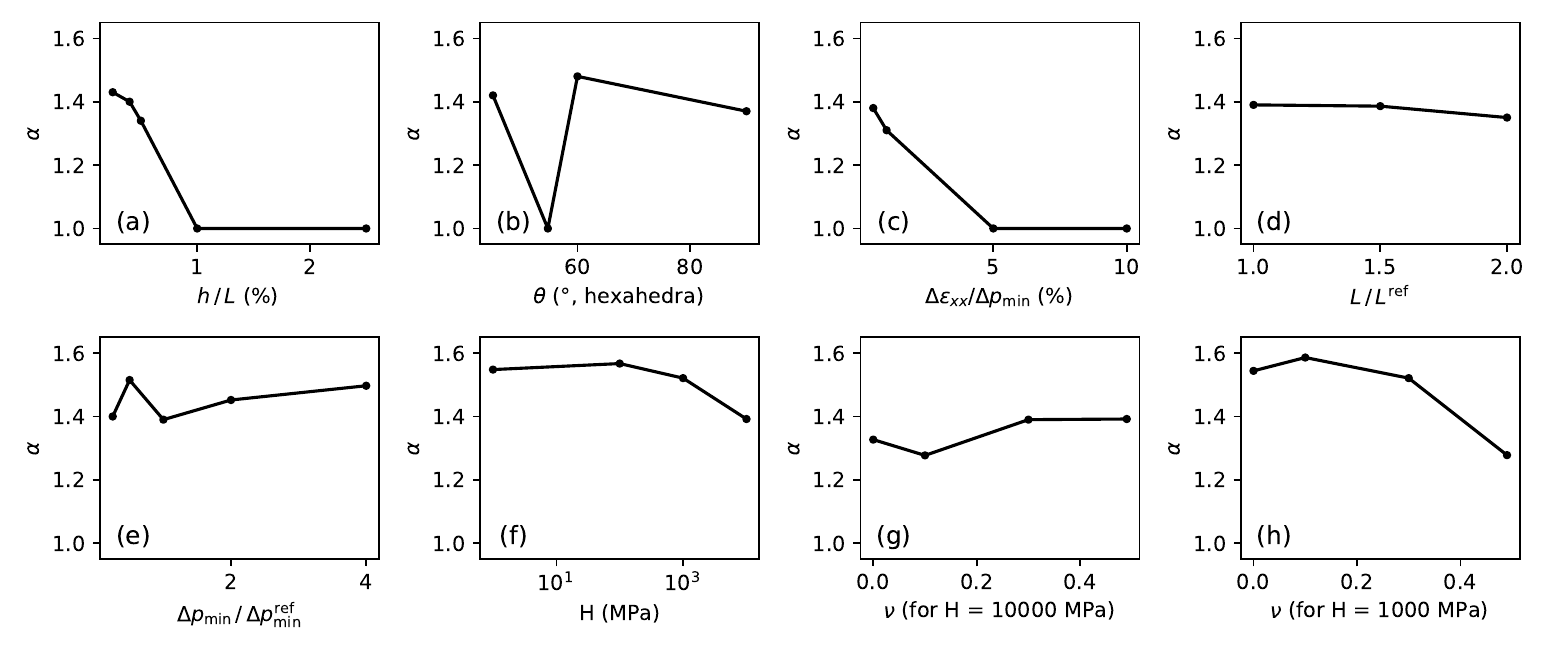}
\caption{Evolution of power law exponent $\alpha$ as a function of (a) $h/L$, (b) $\theta$ (as defined for hexahedral elements), (c) $\Delta \varepsilon_{xx} / \Delta p_\text{min}$, (d) $L/L^{ref}$, (e) $\Delta p_\text{min}/\Delta p_\text{min}^{ref}$, (f) $H$, (g) $\nu$ with $H$ = 10000 MPa and (h) $\nu$ with $H$ = 1000 MPa.}
\label{fig:Effect_On_Alpha}
\end{figure}

The value of the exponent of the power law $\alpha$ is presented for all simulations in Fig. \ref{fig:Effect_On_Alpha}. Fig. \ref{fig:Effect_On_Alpha} (a), (b) and (c) show the values of $\alpha$ for non-model parameters ($h/L$, angle $\theta$ of mesh and $\Delta \varepsilon_{xx} / \Delta p_\text{min}$). The values of $\alpha$ range from 1.0 to 1.5. More precisely, one group of simulations lies around 1.0, while a second group lies between 1.3 and 1.5. The first group corresponds to simulations with a high ratio of $h/L$ and $\Delta \varepsilon_{xx} / \Delta p_\text{min}$, where the simulation distribution did not follow a power law, or with a mesh consisting of hexahedra oriented at $\theta$ = 54.74°, where the plastic events showed atypical and unrepresentative behavior. The second group consists of simulations with finer meshes and prescribed displacements. Simulations with hexahedral meshes that are not aligned too favorably with the ideal orientation of the plastic bands have the same exponent value as simulations with fine enough meshes and small enough displacement steps. It is interesting to note that the power law exponent converges when the mesh is refined and the value of $\Delta \varepsilon_{xx}$ is small enough. It is also noteworthy that the values of the simulation parameters required for convergence are similar to those obtained for the convergence of large stress drop distributions. This implies that small plastic events and large ones are closely related. 

By varying the model parameters, $\alpha$ varies between 1.3 and 1.6 as shown in Fig. \ref{fig:Effect_On_Alpha} (e)-(h). No clear trend can be seen for the effect of $\Delta p_\text{min}$ on $\alpha$, unlike that found for the large stress drop characteristics. When $\Delta p_\text{min}$ is small or large, $\alpha$ is close to 1.45. For $H$, values below 1000~MPa lead to $\alpha$ being close to 1.55, while increasing $H$ to 10000~MPa leads to a decreased value of $\alpha$ = 1.4. When $H$ equals 1000~MPa, having $\nu$ less than 0.3 leads to a value $\alpha$ close to 1.55, while it decreases to 1.3 when $\nu$ = 0.49. When $H$ is 10000~MPa, varying $\nu$ does not affect $\alpha$, which remains between 1.3 and 1.4. Finally, if the value of the sample length is increased from $L_{ref}$ to 2$L_{ref}$, the value of alpha remains around 1.4, which means that the length has no direct effect on $\alpha$.

In the literature on the plasticity of polycrystalline materials, a power law exponent $\alpha$ between 1 and 1.6 is typically observed for tensile deformation, which is consistent with the values predicted by this model~\citep{Csikor2007,Perchikov2024,Weiss2007,Brown2012,Patinet2011,Dimiduk2006}. For example, \citep{Borasi2023} found an exponent $\alpha$ of 1.5 for the tensile deformation of microcast silver crystals with dogbone geometry. This range of exponents is also consistent with those found for metals affected by the PLC effect~\citep{Lebyodkin2000,Ananthakrishna2005}. Although the system-sized plastic events in the model deviate from the experimentally observed behavior due to their Gaussian nature, the small-scale plastic events exhibit a phenomenology similar to that seen in experiments and physical models (MD, DDD, MTM). Thus, the model demonstrates an ability to capture certain key aspects of the underlying physics behind intermittent plasticity.

\subsection{Spatiotemporal analysis of strain bursts}

\begin{figure}[!h]
\centering
\includegraphics[width=1\textwidth]{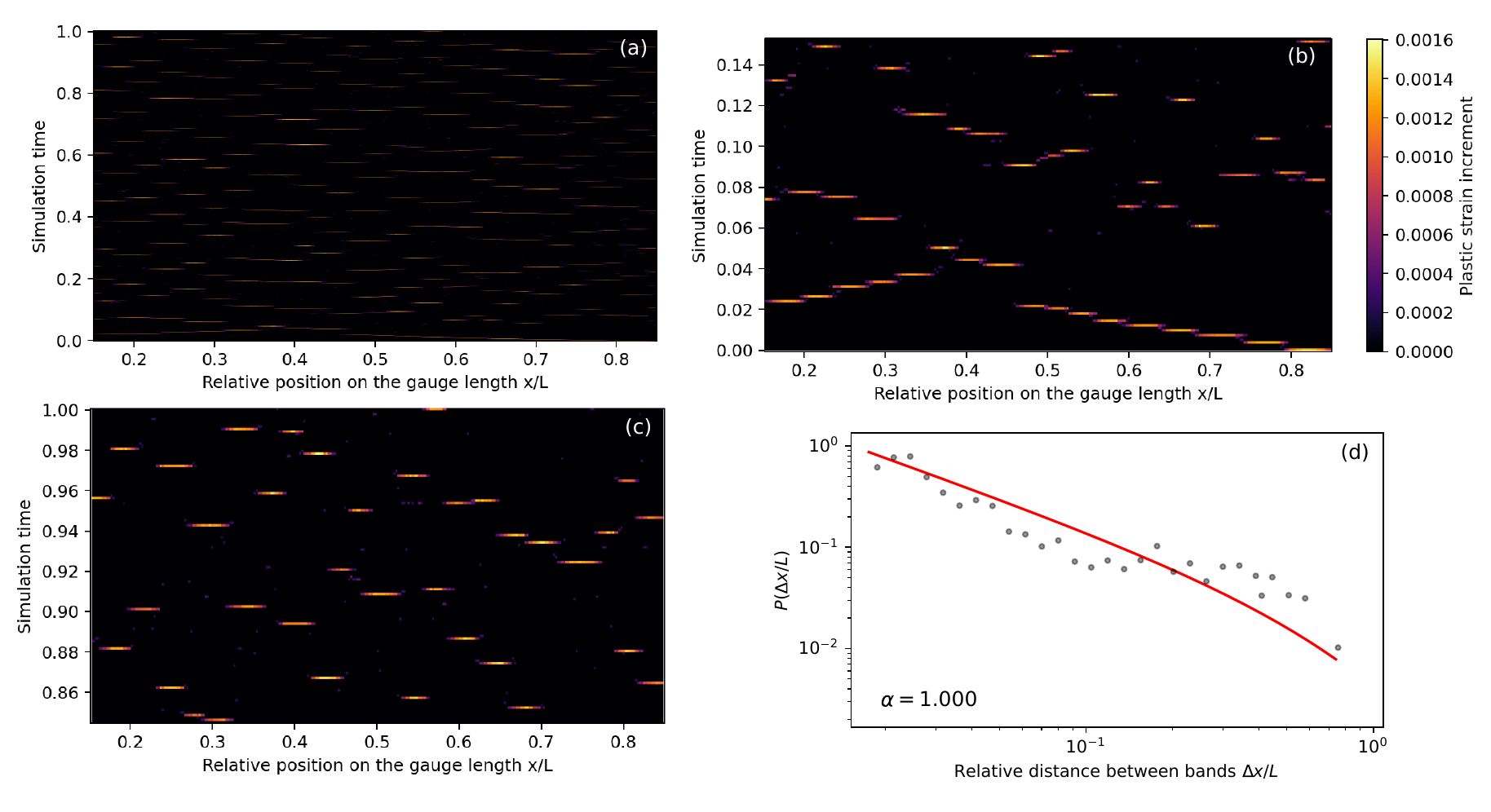}
\caption{(a) Spatiotemporal evolution of plastic strain increment during the reference simulation. X-axis is the relative position on the gauge length (0.15 is the left end of the gauge length, 0.85 its right end). Y-axis is the normalized simulation time. (b) Zoom of (a) at the start of the simulation. (c) Zoom of (a) at the end of the simulation. (d) Distribution of distances between consecutive bands (blue) and the corresponding modeled power law (red). }
\label{fig:Spatiotemporal}
\end{figure}

The spatiotemporal evolution of plastic strain increment on the $O_x$ axis is given in Fig. \ref{fig:Spatiotemporal} for the reference simulation, with a zoom at the start and at the end of the simulation. In those maps, horizontal lines are plastic bands happening in one numerical step. 

The first series of bands, from $t$~=~0 to 0.02, has a "pseudo-propagative" character in the sense that the n-th band will always appear next to the (n-1)-th band. In some simulations, this propagation goes from one end of the gauge to the other. In the current simulation, a second series of "pseudo-propagative" bands appear at the left end of the gauge length instead of only one, at $t$~=~0.021. Then, a third series of correlated bands appear at $t$~=~0.04, as a continuation of the first series. The behavior of band nucleation is phenomenologically analogous to type B PLC~\citep{Yilmaz2011}. However, this first propagation occurs at a stress level whose average value remains constant, which is more similar to a Lüders phenomenon. The reason for this correlated nucleation of bands is the overstress appearing at the boundary of each band, the work-hardening rate $H$ > 0 and the absence of preliminary defect on the gauge length (no concentration of stress prior to the plastic deformation). When $H$ is less than 100 MPa, bands appear but are confined to the area between the gauge length and the head of the specimen where overstress exists due to the dogbone geometry.

Once the entire tensile specimen has been plastically deformed for the first time ($t$~=~0.06), elements with significant local stress concentrations are now randomly distributed throughout the gauge length of the specimen. Between $t$~=~0.06 and 0.15, plastic bands appear more random and less correlated. Between $t$~=~0.85 and 1 (\ref{fig:Spatiotemporal} (c)), bands are nucleating purely randomly on the gauge length, and no series of pseudo-propagative bands are found. This regime of nucleation is comparable to type C PLC~\citep{Yilmaz2011}. 

The distances between consecutive bands (distance between their center) have been measured during the experiment, and their distribution normalized by the specimen length is shown in Fig. \ref{fig:Spatiotemporal} (d) in logarithmic scale. It is found that they are distributed following a power law with an exponent $\alpha$~=~1.0, which is also typical of type C PLC~\citep{Jiang2005}. 

The interest of the presented model is to spontaneously induce macroscopic deformation bands within the specimen. These bands, with finite width, can be interpreted as shear bands caused by an avalanche of dislocations. Ultimately, while remaining within an isotropic elastoplastic framework implementable in finite element codes, the model mimics the same phenomenology as that due to the self-organization of dislocations among themselves in real materials, which is moreover an emerging phenomenon in the model. 
 
\subsection{Application to structural computations}

To evaluate the model under complex multiaxial conditions, simulations were performed on a 3D holed plate geometry. Both the classical continuous plasticity (CCP) and the newly developed time-discontinuous plasticity (TDP) have been tested. Fig.~\ref{fig:HoledPlate} shows the tensile curves of both simulations, the distribution of stress drops from the TDP simulation and the cumulative plastic strain maps for the CCP simulation at 3 given deformation step. The boundary conditions consists of mixed conditions on left and right faces (only axial displacement imposed), the other faces being free. Rigid body motion is suitably fixed. Model parameters used are the same as for the reference dogbone geometry simulation (see Table \ref{tab:params}). Meshes were very refined around the hole ($h/L$ = 0.25\%) and coarser around the edge of the specimen ($h/L$ = 1\%).  The simulation was done in 3D and there were 4 tetrahedra in the thickness.

\begin{figure}[!h]
\centering
\includegraphics[width=\textwidth]{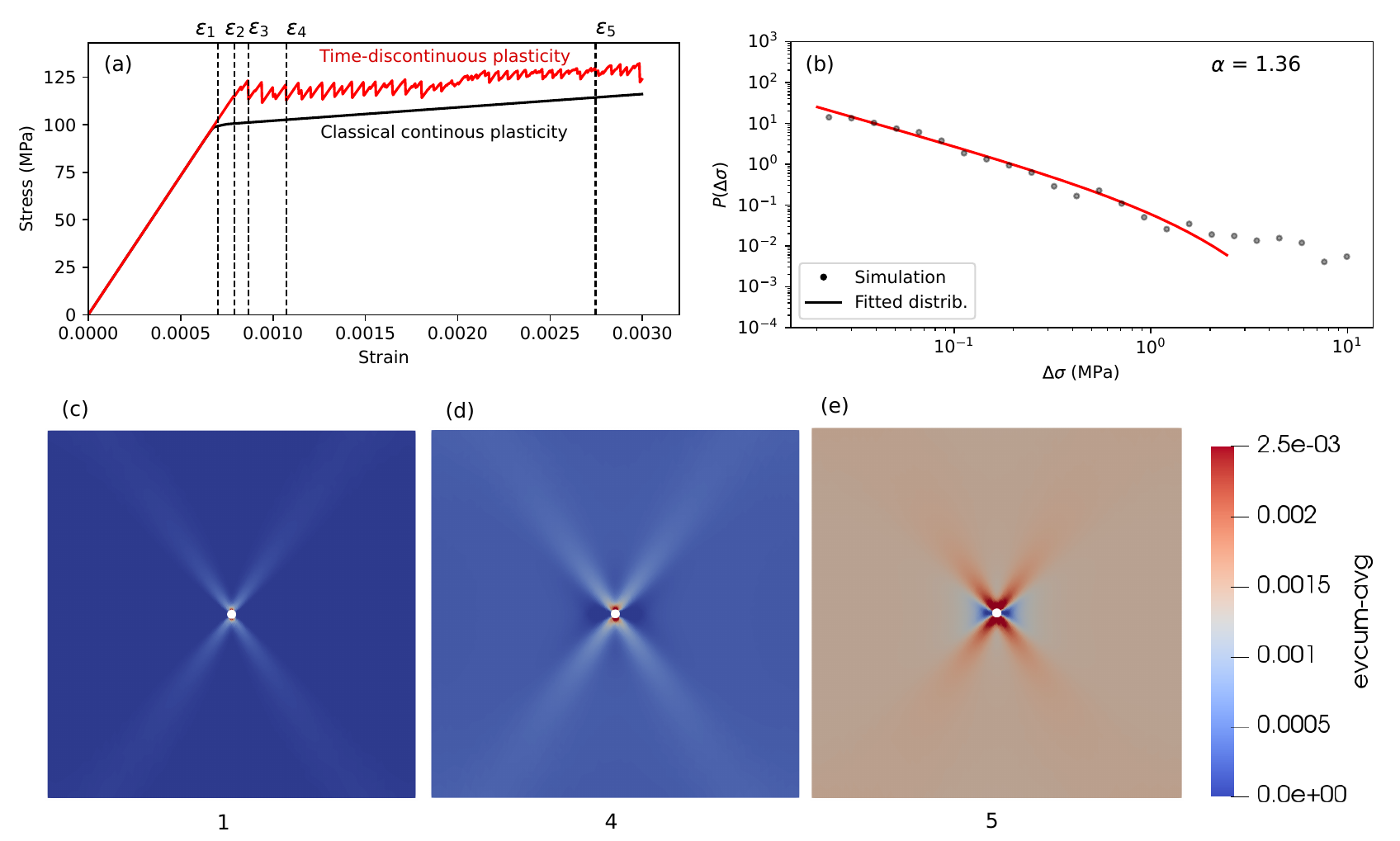}
\caption{(a) Tensile curves obtained from the simulations of a holed plate geometry, using either the classical continuous plasticity (black) or the time-discontinuous plasticity (red). (b) Distribution of stress drops from the time-discontinuous simulation (black dots) and the corresponding power law (red solid line). (c) Cumulative plastic strain maps for the continuous plasticity simulation at $\varepsilon_1$, (d) $\varepsilon_4$ and (e) $\varepsilon_5$.}
\label{fig:PlateHoled1}
\end{figure}

The CCP plate begins to yield macroscopically at $\varepsilon_1$ = 0.06\%, when the average stress within the plate becomes greater than $\sigma_\text{ys}$ 100~MPa, as seen on the corresponding tensile curve at the slope break. Before $\varepsilon_1$ only the area around the hole will yield plastically. After $\varepsilon_1$ the whole plate deforms plastically. The cumulative plastic strain maps show that the plasticity is concentrated around the hole at $\varepsilon_1$, $\varepsilon_4$ and $\varepsilon_5$, forming the x-shape expected for this geometry as it corresponds to the maximum values for von Mises stress.

The tensile curve of the TDP plate shows some serrations, as found for the dogbone geometry. The effective yield stress is 125~MPa, which is higher than the CCP tensile curve, as expected from previous analysis on dogbone specimens. The stress drop distribution (Fig.~\ref{fig:PlateHoled1} (b)) is calculated as indicated in section~\ref{sec:Methodology}, using the axial stress. Although this value does not have the same meaning as before ($\sigma_\text{vM}$ is different from $\sigma_\text{xx}~$ due to the hole), it is still indicative of plastic activity. The shape of the curves is slightly different, with a first part (below 2.5~MPa) following a power law and the second part following a plateau instead of a Gaussian distribution. This is very similar to the supercritical distributions observed for metallic nanopillars smaller than 500 nm in diameter~\citep{ZhangSalman2020}. The first part has been modeled with a truncated power law, defined in Eq. \eqref{DistributionExpression}. The power law exponent $\alpha$ found is 1.36, which is very close to the values found in the dogbone geometry. This suggests that the rate of nucleation of small plastic events at the boundaries of plastic bands is also geometry, and thus intrinsic to the time-discontinuous model.

\begin{figure}[!h]
\centering
\includegraphics[width=0.73\textwidth]{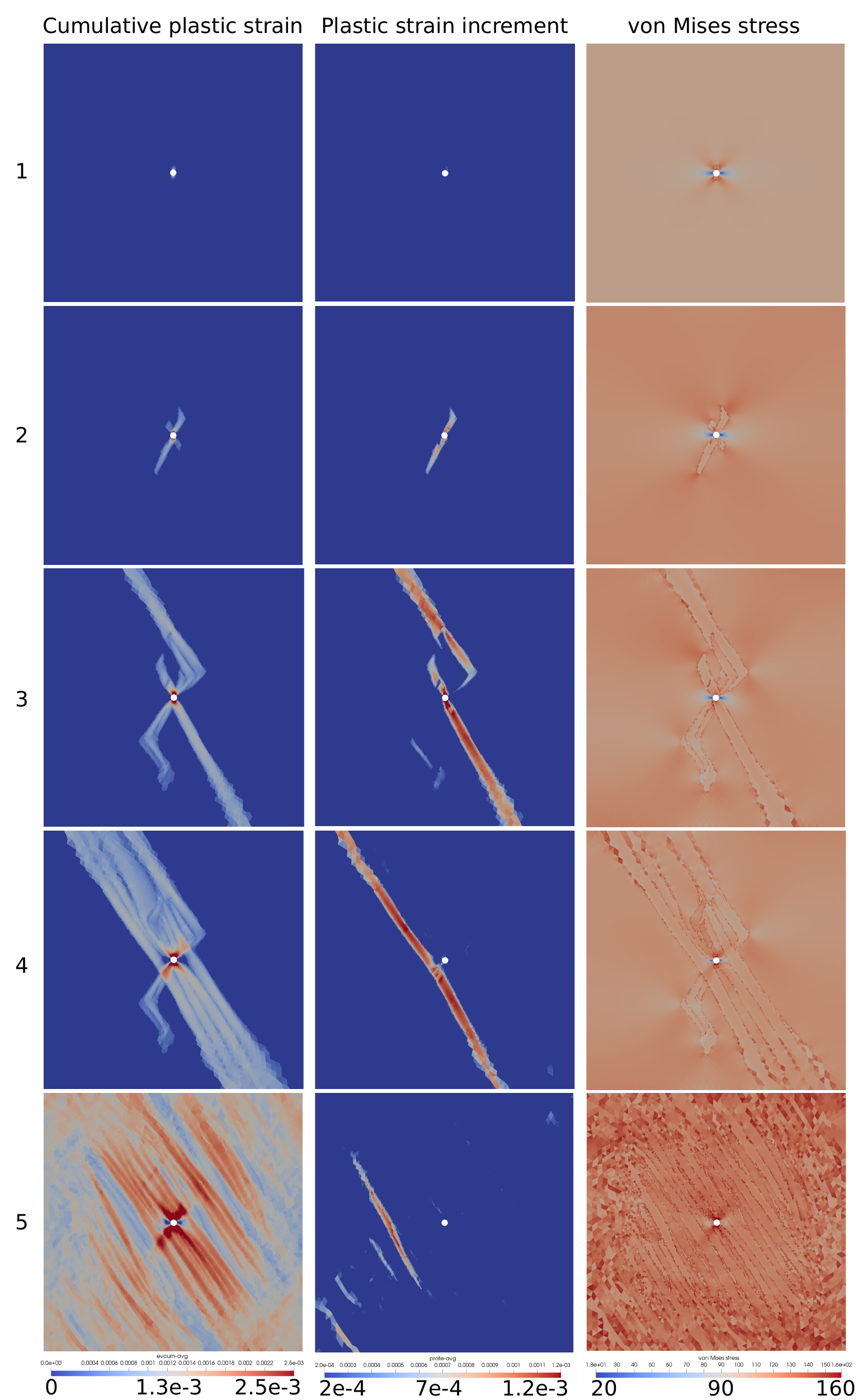}
\caption{Map of cumulative plastic strain, plastic strain increment and von Mises stress at five different levels of deformation of the time-discontinuous simulation. The strain levels are shown in Fig.~\ref{fig:PlateHoled1}. At stage 1, the fields are close to those obtained with classical plasticity. At stage 2 the first significant band appears. At stage 3, the first traversing band appears. At stage 4, several traversing bands have appeared. Stage 5 is the state at the end of the simulation.}
\label{fig:HoledPlate}
\end{figure}

The maps of cumulative plastic strain, plastic strain increment and von Mises stress for the time-discontinuous simulations are displayed in Fig.~\ref{fig:HoledPlate}. As observed in continuous plasticity for this geometry, stress concentration is found around the hole, resulting in higher cumulative plastic strain fields above and below the hole (step 1). In contrast to classical continuum plasticity, where the strain around the hole appears gradually, the plastic strain in our model appears in a spatially and temporally discontinuous manner. 

The first plastic band appearing break the central symmetry of the system (step 2). It is not traversing, with a width of 9 elements. At its edge, von Mises stress concentration is found, which will induce new bands, not necessarily with the same orientation.  

The first traversing plastic bands appear in step 3, associated with the first serration on the tensile curve at $\varepsilon_3$. It is not contiguous. The average plastic strain of the band is the same as for the equivalent dogbone simulation ($\approx$ 5 $\Delta p_\text{min}$). The width of the band increases as it extends away from the central hole. At the boundary of the sample it has a width of 0.04 $L$, which is the average width of the bands in the corresponding dogbone simulation. The following plastic bands appear next to the previous one due to von Mises stress concentration, with the same orientation (step 4).

Similar to the dogbone geometry, plastic activity consists of small plastic events (occurring in some elements, not always contiguous) and large plastic events. The large plastic events can manifest as traversing plastic bands or non-traversing bands around the hole. At the end of the simulation (step 5), the entire sample is covered by plastic bands with the same orientation of $\approx$ 54.74°. This phenomenon can be understood by the shape of the residual von Mises stress, which follows the same direction and favours this direction.

Finally, the comparison of the cumulative plastic strain map at $\varepsilon_5$ of the CCP plate (Fig.~\ref{fig:PlateHoled1} (e)) and the TDP plate (Fig.~\ref{fig:HoledPlate}) reveals a key characteristic of the model. While both plates exhibit large-scale plastic deformation, the map for the time-discontinuous model reveals significant heterogeneity in plastic strain distribution, whereas the CCP map shows a more homogeneous deformation away from the hole. Therefore, this demonstrates that the model successfully achieves the critical objective of obtaining intense localization of plastic deformation into discrete bands, as found in intermittent plasticity~\citep{Charpagne2021}. Unlike previous models, this result is achieved without the introduction of probabilistic elements into the plasticity model itself, thus eliminating the need for prior knowledge of the laws governing plastic flow correlations~\citep{Gelebart2021,Marano2019}.

\section{Conclusions}

A new time-discontinuous plasticity model was developed in this work and implemented independently in the \href{https://docs.fenicsproject.org/}{FEniCSX}~\citep{Scroggs2022} and \href{http://www.zset-software.com/}{Zset}~\citep{Besson1997} solvers. By introducing a single new parameter to conventional J2-plasticity, called the plastic threshold $\Delta p_{\min}$, the model is able to capture the spatial and temporal localization of plastic flow. 

Simulations under homogeneous deformation conditions demonstrate that plastic deformation occurs through abrupt increases in plastic strain, comparable to dislocation avalanches, and these events are always associated with sharp stress drops. The model is interpreted through the existence of two yield surfaces: the upper surface controls the initiation of plasticity, while the lower surface defines the stress reached after plastic relaxation. The results of the new model under homogeneous fields are fully understood under any triaxiality.

When the geometry of the specimen is complex and induces stress concentration, plastic strain bands spontaneously emerge in a random manner, without introducing any stochasticity as an input to the model. Stochasticity of plastic events is an outcome of the model, not an input in contrast to many existing models in this field. Temporal intermittency of plastic activity and consecutive stress drops obey a power law for small events and a Gaussian distribution for large events. Spatial intermittency of plastic activity follows a power law. The effect of model and simulation parameters, such as $\Delta p_{\min}$, elastic constants, work-hardening modulus and mesh size, have been extensively study. Notably the band widths are proportional to the length of the tensile specimen, and the plastic strain carried by the band is proportional to $\Delta p_{\min}$. The model is homothetic in nature as it contains no internal length. It has been proven to be usable under any geometry, no matter its complexity, while remaining efficient in terms of calculation time. {\color{black} One limitation of the model is that, to obtain serrations and reliable statistics on stress drops and plastic events, the applied strain in the specimen must be at least below to 5\% of $\Delta p_\text{min}$.} No spurious mesh dependency is observed in contrast to softening models for localization. 

This model developed in the von Mises plasticity framework is the first step towards the development of a time-discontinuous model in crystal plasticity. It is believed that this next step is necessary for the current work to be more physically relevant and thus comparable to experimental data. To control band size, a statistical spatial distribution of $\Delta p_{\min}$ across the specimen should be introduced into the model. The statistic of $\Delta p_{\min}$ could be calibrated using data from MD, MTM, or DDD models, which represent the most physically relevant approaches.

\section*{Declaration of Competing Interest}

The authors declare that they have no known competing financial interests or personal relationships that could have appeared to influence the work reported in this paper.

\section*{Funding sources}

This work was funded by the French ANR (Agence Nationale de la Recherche) under the MESOCRYSP project (ANR-21-CE08-0030).

\section*{Declaration of generative AI in scientific writing}

During the preparation of this work the author(s) used ChatGPT 3.5 and DeepL in order to improve language and readability. After using this tool/service, the author(s) reviewed and edited the content as needed and take(s) full responsibility for the content of the published article.

\section*{Data Availability Statement}

The data that support the findings of this study are openly available in the Zenodo repository titled 'Time-discontinuous plasticity, data and videos' at https://zenodo.org/records/14266823, reference number 10.5281/zenodo.14266823. The code used for FEniCSX simulations can be found at https://github.com/Mathias-Lamari/Time-discontinuous-plasticity.

\bibliographystyle{plainnat}
\bibliography{cas-refs}

\begin{thebibliography}{58}
\providecommand{\natexlab}[1]{#1}
\providecommand{\url}[1]{\texttt{#1}}
\expandafter\ifx\csname urlstyle\endcsname\relax
  \providecommand{\doi}[1]{doi: #1}\else
  \providecommand{\doi}{doi: \begingroup \urlstyle{rm}\Url}\fi

\bibitem[Alstott et~al.(2014)Alstott, Bullmore, and Plenz]{Alstott2014}
J.~Alstott, E.~Bullmore, and D.~Plenz.
\newblock powerlaw: A python package for analysis of heavy-tailed
  distributions.
\newblock \emph{PLoS ONE}, 9\penalty0 (1):\penalty0 e85777, January 2014.
\newblock ISSN 1932-6203.
\newblock \doi{10.1371/journal.pone.0085777}.
\newblock URL \url{http://dx.doi.org/10.1371/journal.pone.0085777}.

\bibitem[Ananthakrishna(2005)]{Ananthakrishna2005}
G.~Ananthakrishna.
\newblock Spatio-temporal features of the portevin-le chatelier effect.
\newblock \emph{Materials Science and Engineering: A}, 400-401:\penalty0
  210--213, 2005.
\newblock ISSN 0921-5093.
\newblock \doi{https://doi.org/10.1016/j.msea.2005.03.037}.
\newblock URL
  \url{https://www.sciencedirect.com/science/article/pii/S0921509305002650}.
\newblock Dislocations 2004.

\bibitem[Baggio et~al.(2023{\natexlab{a}})Baggio, Salman, and
  Truskinovsky]{Baggio2023}
R.~Baggio, O.~U. Salman, and L.~Truskinovsky.
\newblock Inelastic rotations and pseudoturbulent plastic avalanches in
  crystals.
\newblock \emph{Physical Review E}, 107:\penalty0 005000--1--005000--19,
  2023{\natexlab{a}}.

\bibitem[Baggio et~al.(2023{\natexlab{b}})Baggio, Salman, and
  Truskinovsky]{Baggio2023bis}
R.~Baggio, O.~U. Salman, and L.~Truskinovsky.
\newblock Homogeneous nucleation of dislocations as a pattern formation
  phenomenon.
\newblock \emph{European Journal of Mechanics - A/Solids}, 99:\penalty0 104897,
  2023{\natexlab{b}}.
\newblock ISSN 0997-7538.
\newblock \doi{https://doi.org/10.1016/j.euromechsol.2022.104897}.
\newblock URL
  \url{https://www.sciencedirect.com/science/article/pii/S0997753822003278}.

\bibitem[Besson and Foerch(1997)]{Besson1997}
J.~Besson and R.~Foerch.
\newblock Large scale object-oriented finite element code design.
\newblock \emph{Computer Methods in Applied Mechanics and Engineering},
  142:\penalty0 165--187, 1997.

\bibitem[Besson et~al.(2009)Besson, Cailletaud, Chaboche, and
  Forest]{Besson2009}
J.~Besson, G.~Cailletaud, J.-L. Chaboche, and S.~Forest.
\newblock \emph{Non-linear mechanics of materials}, volume 167.
\newblock Springer Science \& Business Media, 2009.

\bibitem[Borasi et~al.(2023)Borasi, Frasca, Charbon, and Mortensen]{Borasi2023}
L.~Borasi, S.~Frasca, E.~Charbon, and A.~Mortensen.
\newblock The effect of size, orientation and temperature on the deformation of
  microcast silver crystals.
\newblock \emph{Acta Materialia}, 249:\penalty0 118817, 2023.
\newblock ISSN 1359-6454.
\newblock \doi{https://doi.org/10.1016/j.actamat.2023.118817}.
\newblock URL
  \url{https://www.sciencedirect.com/science/article/pii/S1359645423001489}.

\bibitem[Brown(2012)]{Brown2012}
L.~M. Brown.
\newblock Constant intermittent flow of dislocations: central problems in
  plasticity.
\newblock \emph{Materials Science and Technology}, 28\penalty0 (11):\penalty0
  1209--1232, 2012.
\newblock \doi{10.1179/174328412X13409726212768}.
\newblock URL \url{https://doi.org/10.1179/174328412X13409726212768}.

\bibitem[Charpagne et~al.(2021)Charpagne, Hestroffer, Polonsky, Echlin, Texier,
  Valle, Beyerlein, Pollock, and Stinville]{Charpagne2021}
M.~A. Charpagne, J.~M. Hestroffer, A.~T. Polonsky, M.~P. Echlin, D.~Texier,
  V.~Valle, I.~J. Beyerlein, T.~M. Pollock, and J.~C. Stinville.
\newblock Slip localization in inconel 718: A three-dimensional and statistical
  perspective.
\newblock \emph{Acta Materialia}, 215:\penalty0 117037, 2021.
\newblock ISSN 1359-6454.
\newblock \doi{https://doi.org/10.1016/j.actamat.2021.117037}.
\newblock URL
  \url{https://www.sciencedirect.com/science/article/pii/S1359645421004171}.

\bibitem[Chevalier et~al.(2018)Chevalier, Brassart, Lani, Bailly, Pardoen, and
  Morelle]{Chevalier2018}
J.~Chevalier, L.~Brassart, F.~Lani, C.~Bailly, T.~Pardoen, and X.~P. Morelle.
\newblock Unveiling the nanoscale heterogeneity controlled deformation of
  thermosets.
\newblock \emph{Journal of the Mechanics and Physics of Solids}, 121:\penalty0
  432--446, 2018.
\newblock ISSN 0022-5096.
\newblock \doi{https://doi.org/10.1016/j.jmps.2018.08.014}.
\newblock URL
  \url{https://www.sciencedirect.com/science/article/pii/S0022509618303909}.

\bibitem[Clauset et~al.(2009)Clauset, Shalizi, and Newman]{Aaron2009}
A.~Clauset, C.~R. Shalizi, and M.~E.~J. Newman.
\newblock Power-law distributions in empirical data.
\newblock \emph{SIAM Review}, 51\penalty0 (4):\penalty0 661--703, 2009.
\newblock \doi{10.1137/070710111}.
\newblock URL \url{https://doi.org/10.1137/070710111}.

\bibitem[Colas et~al.(2014)Colas, Finot, Flouriot, Forest, Mazière, and
  Paris]{Colas2014}
D.~Colas, E.~Finot, S.~Flouriot, S.~Forest, M.~Mazière, and T.~Paris.
\newblock Investigation and modeling of the anomalous yield point phenomenon in
  pure tantalum.
\newblock \emph{Materials Science and Engineering: A}, 615:\penalty0 283--295,
  2014.
\newblock ISSN 0921-5093.
\newblock \doi{https://doi.org/10.1016/j.msea.2014.07.028}.
\newblock URL
  \url{https://www.sciencedirect.com/science/article/pii/S0921509314008934}.

\bibitem[Csikor et~al.(2007)Csikor, Motz, Weygand, Zaiser, and
  Zapperi]{Csikor2007}
F.~F. Csikor, C.~Motz, D.~Weygand, M.~Zaiser, and S.~Zapperi.
\newblock Dislocation avalanches, strain bursts, and the problem of plastic
  forming at the micrometer scale.
\newblock \emph{Science}, 318\penalty0 (5848):\penalty0 251--254, 2007.
\newblock \doi{10.1126/science.1143719}.
\newblock URL \url{https://www.science.org/doi/abs/10.1126/science.1143719}.

\bibitem[de~Geus and Wyart(2022)]{Geus2022}
T.~W.~J. de~Geus and M.~Wyart.
\newblock Scaling theory for the statistics of slip at frictional interfaces.
\newblock \emph{Phys. Rev. E}, 106:\penalty0 065001, Dec 2022.
\newblock \doi{10.1103/PhysRevE.106.065001}.
\newblock URL \url{https://link.aps.org/doi/10.1103/PhysRevE.106.065001}.

\bibitem[de~Souza~Neto et~al.(2011)de~Souza~Neto, Peric, and Owen]{Souza2011}
E.~A. de~Souza~Neto, D.~Peric, and D.~R.~J. Owen.
\newblock \emph{Computational methods for plasticity: theory and applications}.
\newblock John Wiley \& Sons, 2011.

\bibitem[Dimiduk et~al.(2006)Dimiduk, Woodward, LeSar, and Uchic]{Dimiduk2006}
D.~M. Dimiduk, C.~Woodward, R.~LeSar, and M.~D. Uchic.
\newblock Scale-free intermittent flow in crystal plasticity.
\newblock \emph{Science}, 312\penalty0 (5777):\penalty0 1188--1190, 2006.
\newblock ISSN 00368075, 10959203.
\newblock URL \url{http://www.jstor.org/stable/3846259}.

\bibitem[Fernández-Castellanos et~al.(2021)Fernández-Castellanos, Roux, and
  Patinet]{Castellanos2021}
D.~Fernández-Castellanos, S.~Roux, and S.~Patinet.
\newblock Insights from the quantitative calibration of an elasto-plastic model
  from a lennard-jones atomic glass.
\newblock \emph{Comptes Rendus Physique}, 22:\penalty0 135--162, 2021.

\bibitem[G\'el\'ebart(2021)]{Gelebart2021}
L.~G\'el\'ebart.
\newblock Grain size effects and weakest link theory {in~3D} crystal plasticity
  simulations of polycrystals.
\newblock \emph{Comptes Rendus. Physique}, 22\penalty0 (S3):\penalty0 313--330,
  2021.
\newblock \doi{10.5802/crphys.53}.

\bibitem[Greer et~al.(2013)Greer, Cheng, and Ma]{Greer2013}
A.~L. Greer, Y.~Q. Cheng, and E.~Ma.
\newblock Shear bands in metallic glasses.
\newblock \emph{Materials Science and Engineering: R: Reports}, 74\penalty0
  (4):\penalty0 71--132, 2013.
\newblock ISSN 0927-796X.
\newblock \doi{https://doi.org/10.1016/j.mser.2013.04.001}.
\newblock URL
  \url{https://www.sciencedirect.com/science/article/pii/S0927796X13000259}.

\bibitem[Guillermin et~al.(2023)Guillermin, Besson, Koster, Lacourt, Maziere,
  Chalons, and Forest]{Guillermin2023}
N.~Guillermin, J.~Besson, A.~Koster, L.~Lacourt, M.~Maziere, H.~Chalons, and
  S.~Forest.
\newblock Experimental and numerical analysis of the portevin–le chatelier
  effect in a nickel-base superalloy for turbine disks application.
\newblock \emph{International Journal of Solids and Structures}, 264:\penalty0
  112076, 2023.
\newblock ISSN 0020-7683.
\newblock \doi{https://doi.org/10.1016/j.ijsolstr.2022.112076}.
\newblock URL
  \url{https://www.sciencedirect.com/science/article/pii/S0020768322005297}.

\bibitem[Gómez-García et~al.(2006)Gómez-García, Devincre, and
  Kubin]{Gomez-Garcia2006}
D.~Gómez-García, B.~Devincre, and L.~P. Kubin.
\newblock Dislocation patterns and the similitude principle: 2.5d mesoscale
  simulations.
\newblock \emph{Physical Review Letters}, 96:\penalty0 125503--1--125503--6,
  2006.

\bibitem[Jiang et~al.(2005)Jiang, Zhang, Jiang, Chen, and Wu]{Jiang2005}
Z.~Jiang, Q.~Zhang, H.~Jiang, Z.~Chen, and X.~Wu.
\newblock Spatial characteristics of the portevin-le chatelier deformation
  bands in al-4at
\newblock \emph{Materials Science and Engineering: A}, 403\penalty0
  (1):\penalty0 154--164, 2005.
\newblock ISSN 0921-5093.
\newblock \doi{https://doi.org/10.1016/j.msea.2005.05.059}.
\newblock URL
  \url{https://www.sciencedirect.com/science/article/pii/S0921509305004934}.

\bibitem[Kerfriden(2024)]{Kerfriden_2023}
P.~Kerfriden.
\newblock Quantized plasticity 10cm, pierre kerfriden.
\newblock \url{https://www.youtube.com/watch?v=nW0KEG-W4jk}, 2024.

\bibitem[Kiener et~al.(2008)Kiener, Grosinger, Dehm, and Pippan]{Kiener2008}
D.~Kiener, W.~Grosinger, G.~Dehm, and R.~Pippan.
\newblock A further step towards an understanding of size-dependent crystal
  plasticity: In situ tension experiments of miniaturized single-crystal copper
  samples.
\newblock \emph{Acta Materialia}, 56\penalty0 (3):\penalty0 580--592, 2008.
\newblock ISSN 1359-6454.
\newblock \doi{https://doi.org/10.1016/j.actamat.2007.10.015}.
\newblock URL
  \url{https://www.sciencedirect.com/science/article/pii/S1359645407006969}.

\bibitem[Lamari et~al.(2024)Lamari, Allain, Geandier, Ponçot, Perlade, and
  Zhu]{Lamari2024}
M.~Lamari, S.~Y.~P. Allain, G.~Geandier, M.~Ponçot, A.~Perlade, and K.~Zhu.
\newblock Behavior of trip-aided medium mn steels investigated by in situ
  synchrotron x-ray diffraction experiments and microstructure-based
  micromechanical modelling.
\newblock \emph{International Journal of Plasticity}, 173:\penalty0 103866,
  2024.
\newblock ISSN 0749-6419.
\newblock \doi{https://doi.org/10.1016/j.ijplas.2023.103866}.
\newblock URL
  \url{https://www.sciencedirect.com/science/article/pii/S0749641923003509}.

\bibitem[Lebyodkin et~al.(2000)Lebyodkin, Dunin-Barkowskii, Bréchet, Estrin,
  and Kubin]{Lebyodkin2000}
M.~Lebyodkin, L.~Dunin-Barkowskii, Y.~Bréchet, Y.~Estrin, and L.P. Kubin.
\newblock Spatio-temporal dynamics of the portevin–le chatelier effect:
  experiment and modelling.
\newblock \emph{Acta Materialia}, 48\penalty0 (10):\penalty0 2529--2541, 2000.
\newblock ISSN 1359-6454.
\newblock \doi{https://doi.org/10.1016/S1359-6454(00)00067-7}.
\newblock URL
  \url{https://www.sciencedirect.com/science/article/pii/S1359645400000677}.

\bibitem[Marano et~al.(2019)Marano, Gelebart, and Forest]{Marano2019}
A.~Marano, L.~Gelebart, and S.~Forest.
\newblock Intragranular localization induced by softening crystal plasticity:
  Analysis of slip and kink bands localization modes from high resolution
  fft-simulations results.
\newblock \emph{Acta Materialia}, 175:\penalty0 262--275, 2019.
\newblock ISSN 1359-6454.
\newblock \doi{https://doi.org/10.1016/j.actamat.2019.06.010}.
\newblock URL
  \url{https://www.sciencedirect.com/science/article/pii/S1359645419303696}.

\bibitem[Mazi{\`e}re and Forest(2015)]{Maziere2015}
M.~Mazi{\`e}re and S.~Forest.
\newblock Strain gradient plasticity modeling and finite element simulation of
  l{\"u}ders band formation and propagation.
\newblock \emph{Continuum Mechanics and Thermodynamics}, 27\penalty0
  (1):\penalty0 83--104, 2015.
\newblock ISSN 1432-0959.
\newblock \doi{10.1007/s00161-013-0331-8}.
\newblock URL \url{https://doi.org/10.1007/s00161-013-0331-8}.

\bibitem[Oh et~al.(2009)Oh, Legros, Kiener, and Dehm]{Oh2009}
S.~H. Oh, M.~Legros, D.~Kiener, and G.~Dehm.
\newblock In situ observation of dislocation nucleation and escape in a
  submicrometre aluminium single crystal.
\newblock \emph{Nature Materials}, 8\penalty0 (2):\penalty0 95--100, 02 2009.
\newblock ISSN 1476-4660.
\newblock \doi{10.1038/nmat2370}.
\newblock URL \url{https://doi.org/10.1038/nmat2370}.

\bibitem[Okuyucu et~al.(2023)Okuyucu, Ulucan, Abboud, Motallebzadeh, Özerinç,
  Kalay, and Kalay]{Okuyucu2023}
C.~Okuyucu, T.~H. Ulucan, M.~Abboud, A.~Motallebzadeh, S.~Özerinç, İ. Kalay,
  and Y.~E. Kalay.
\newblock Nanomechanical properties of al-tb marginal metallic glass.
\newblock \emph{Materials Science and Engineering: A}, 888:\penalty0 145809,
  2023.
\newblock ISSN 0921-5093.
\newblock \doi{https://doi.org/10.1016/j.msea.2023.145809}.
\newblock URL
  \url{https://www.sciencedirect.com/science/article/pii/S0921509323012339}.

\bibitem[Patinet et~al.(2011)Patinet, Bonamy, and Proville]{Patinet2011}
S.~Patinet, D.~Bonamy, and L.~Proville.
\newblock Atomic-scale avalanche along a dislocation in a random alloy.
\newblock \emph{Phys. Rev. B}, 84:\penalty0 174101, Nov 2011.
\newblock \doi{10.1103/PhysRevB.84.174101}.
\newblock URL \url{https://link.aps.org/doi/10.1103/PhysRevB.84.174101}.

\bibitem[Perchikov and Truskinovsky(2024)]{Perchikov2024}
N.~Perchikov and L.~Truskinovsky.
\newblock Quantized plastic deformation.
\newblock \emph{Journal of the Mechanics and Physics of Solids}, 190:\penalty0
  105704, 2024.
\newblock ISSN 0022-5096.
\newblock \doi{https://doi.org/10.1016/j.jmps.2024.105704}.
\newblock URL
  \url{https://www.sciencedirect.com/science/article/pii/S0022509624001704}.

\bibitem[Perfilyev et~al.(2013)Perfilyev, Moshkovich, Lapsker, Laikhtman, and
  Rapoport]{Perfilyev2013}
V.~Perfilyev, A.~Moshkovich, I.~Lapsker, A.~Laikhtman, and L.~Rapoport.
\newblock The effect of vanadium content and temperature on stick–slip
  phenomena under friction of crv(x)n coatings.
\newblock \emph{Wear}, 307\penalty0 (1):\penalty0 44--51, 2013.
\newblock ISSN 0043-1648.
\newblock \doi{https://doi.org/10.1016/j.wear.2013.08.012}.
\newblock URL
  \url{https://www.sciencedirect.com/science/article/pii/S0043164813004651}.

\bibitem[Richeton et~al.(2005)Richeton, Weiss, and Louchet]{Richeton2005}
T.~Richeton, J.~Weiss, and F.~Louchet.
\newblock Breakdown of avalanche critical behaviour in polycrystalline
  plasticity.
\newblock \emph{Nature Materials}, 4\penalty0 (6):\penalty0 465--469, Jun 2005.
\newblock ISSN 1476-4660.
\newblock \doi{10.1038/nmat1393}.
\newblock URL \url{https://doi.org/10.1038/nmat1393}.

\bibitem[Ruestes and Segurado(2024)]{Ruestes2024}
C.~J. Ruestes and J.~Segurado.
\newblock A stochastic discrete slip approach to microplasticity: Application
  to submicron w pillars.
\newblock \emph{International Journal of Plasticity}, 176:\penalty0 103965,
  2024.
\newblock ISSN 0749-6419.
\newblock \doi{https://doi.org/10.1016/j.ijplas.2024.103965}.
\newblock URL
  \url{https://www.sciencedirect.com/science/article/pii/S0749641924000925}.

\bibitem[Ry{\'{s}} et~al.(2024)Ry{\'{s}}, Kursa, and Petryk]{Rys2024}
M.~Ry{\'{s}}, M.~Kursa, and H.~Petryk.
\newblock Spontaneous emergence of deformation bands in single-crystal
  plasticity simulations at small strain.
\newblock \emph{Computational Mechanics}, Jul 2024.
\newblock ISSN 1432-0924.
\newblock \doi{10.1007/s00466-024-02519-8}.
\newblock URL \url{https://doi.org/10.1007/s00466-024-02519-8}.

\bibitem[Salman and Truskinovsky(2011)]{Salman2011}
O.~U. Salman and L.~Truskinovsky.
\newblock Minimal integer automaton behind crystal plasticity.
\newblock \emph{Phys. Rev. Lett.}, 106:\penalty0 175503, Apr 2011.
\newblock \doi{10.1103/PhysRevLett.106.175503}.
\newblock URL \url{https://link.aps.org/doi/10.1103/PhysRevLett.106.175503}.

\bibitem[Salman and Truskinovsky(2012)]{Salman2012}
O.~U. Salman and L.~Truskinovsky.
\newblock On the critical nature of plastic flow: One and two dimensional
  models.
\newblock \emph{International Journal of Engineering Science}, 59:\penalty0
  219--254, 2012.
\newblock ISSN 0020-7225.
\newblock \doi{https://doi.org/10.1016/j.ijengsci.2012.03.012}.
\newblock URL
  \url{https://www.sciencedirect.com/science/article/pii/S0020722512000559}.
\newblock The Special Issue in honor of VICTOR L. BERDICHEVSKY.

\bibitem[Scroggs et~al.(2022)Scroggs, Dokken, Richardson, and
  Wells]{Scroggs2022}
M.~W. Scroggs, J.~S. Dokken, C.~N. Richardson, and G.~N. Wells.
\newblock Construction of arbitrary order finite element degree-of-freedom maps
  on polygonal and polyhedral cell meshes.
\newblock \emph{ACM Transactions on Mathematical Software}, 48:\penalty0
  18--1--18--23, 2022.

\bibitem[Simo(1998)]{Simo1998}
J.~C. Simo.
\newblock Numerical analysis and simulation of plasticity.
\newblock In \emph{Numerical Methods for Solids (Part 3) Numerical Methods for
  Fluids (Part 1)}, volume~6 of \emph{Handbook of Numerical Analysis}, pages
  183--499. Elsevier, 1998.
\newblock \doi{https://doi.org/10.1016/S1570-8659(98)80009-4}.
\newblock URL
  \url{https://www.sciencedirect.com/science/article/pii/S1570865998800094}.

\bibitem[Sornette and Ouillon(2012)]{Sornette2012}
D.~Sornette and G.~Ouillon.
\newblock Dragon-kings: Mechanisms, statistical methods and empirical evidence.
\newblock \emph{The European Physical Journal Special Topics}, 205\penalty0
  (1):\penalty0 1--26, May 2012.
\newblock ISSN 1951-6401.
\newblock \doi{10.1140/epjst/e2012-01559-5}.
\newblock URL \url{https://doi.org/10.1140/epjst/e2012-01559-5}.

\bibitem[Uchic et~al.(2009)Uchic, Shade, and Dimiduk]{Uchic2009}
M.~D. Uchic, P.~A. Shade, and D.~M. Dimiduk.
\newblock Micro-compression testing of fcc metals: A selected overview of
  experiments and simulations.
\newblock \emph{JOM}, 61\penalty0 (3):\penalty0 36--41, Mar 2009.
\newblock ISSN 1543-1851.
\newblock \doi{10.1007/s11837-009-0038-2}.
\newblock URL \url{https://doi.org/10.1007/s11837-009-0038-2}.

\bibitem[Vermeij et~al.(2024)Vermeij, Wijnen, Peerlings, Geers, and
  Hoefnagels]{Vermeij2024}
T.~Vermeij, J.~Wijnen, R.~H.~J. Peerlings, M.~G.~D. Geers, and J.~P.~M.
  Hoefnagels.
\newblock A quasi-2d integrated experimental–numerical approach to
  high-fidelity mechanical analysis of metallic microstructures.
\newblock \emph{Acta Materialia}, 264:\penalty0 119551, 2024.
\newblock ISSN 1359-6454.
\newblock \doi{https://doi.org/10.1016/j.actamat.2023.119551}.
\newblock URL
  \url{https://www.sciencedirect.com/science/article/pii/S1359645423008807}.

\bibitem[Wang et~al.(2011)Wang, Berdin, Maziere, Forest, Prioul, Parrot, and
  Le-Delliou]{WangMaziere2011}
H.~Wang, C.~Berdin, M.~Maziere, S.~Forest, C.~Prioul, A.~Parrot, and
  P.~Le-Delliou.
\newblock Portevin–le chatelier (plc) instabilities and slant fracture in
  c–mn steel round tensile specimens.
\newblock \emph{Scripta Materialia}, 64\penalty0 (5):\penalty0 430--433, 2011.
\newblock ISSN 1359-6462.
\newblock \doi{https://doi.org/10.1016/j.scriptamat.2010.11.005}.
\newblock URL
  \url{https://www.sciencedirect.com/science/article/pii/S1359646210007530}.

\bibitem[Weiss et~al.(2007)Weiss, Richeton, Louchet, Chmelik, Dobron,
  Entemeyer, Lebyodkin, Lebedkina, Fressengeas, and McDonald]{Weiss2007}
J.~Weiss, T.~Richeton, F.~Louchet, F.~Chmelik, P.~Dobron, D.~Entemeyer,
  M.~Lebyodkin, T.~Lebedkina, C.~Fressengeas, and R.~J. McDonald.
\newblock Evidence for universal intermittent crystal plasticity from acoustic
  emission and high-resolution extensometry experiments.
\newblock \emph{Physical Review B}, 76:\penalty0 224110--1--224110--8, 2007.

\bibitem[Weiss et~al.(2015)Weiss, Rhouma, Richeton, Dechanel, Louchet, and
  Truskinovsky]{Weiss2015}
J.~Weiss, W.~Ben Rhouma, T.~Richeton, S.~Dechanel, F.~Louchet, and
  L.~Truskinovsky.
\newblock From mild to wild fluctuations in crystal plasticity.
\newblock \emph{Physical Review Letters}, 114:\penalty0 105504--1--105504--6,
  2015.

\bibitem[Wijnen et~al.(2021)Wijnen, Peerlings, Hoefnagels, and
  Geers]{Wijnen2021}
J.~Wijnen, R.~H.~J. Peerlings, J.~P.~M. Hoefnagels, and M.~G.~D. Geers.
\newblock A discrete slip plane model for simulating heterogeneous plastic
  deformation in single crystals.
\newblock \emph{International Journal of Solids and Structures}, 228:\penalty0
  111094--1--111094--12, 2021.

\bibitem[Xu et~al.(2022{\natexlab{a}})Xu, Zhao, Wang, Duan, Lei, and
  Fang]{XuZhao2022}
M.~Xu, Z.~Zhao, P.~Wang, S.~Duan, H.~Lei, and D.~Fang.
\newblock Mechanical performance of bio-inspired hierarchical honeycomb
  metamaterials.
\newblock \emph{International Journal of Solids and Structures},
  254-255:\penalty0 111866, 2022{\natexlab{a}}.
\newblock ISSN 0020-7683.
\newblock \doi{https://doi.org/10.1016/j.ijsolstr.2022.111866}.
\newblock URL
  \url{https://www.sciencedirect.com/science/article/pii/S0020768322003341}.

\bibitem[Xu et~al.(2022{\natexlab{b}})Xu, Cheng, Zhao, Wang, and Chen]{Xu2022}
X.~Xu, L.~Cheng, X.~Zhao, J.~Wang, and X.~Chen.
\newblock Micro/nano periodic surface structures and performance of stainless
  steel machined using femtosecond lasers.
\newblock \emph{Micromachines}, 13\penalty0 (6), 2022{\natexlab{b}}.
\newblock ISSN 2072-666X.
\newblock \doi{10.3390/mi13060976}.
\newblock URL \url{https://www.mdpi.com/2072-666X/13/6/976}.

\bibitem[Yan et~al.(2010)Yan, Yan, Hu, and Dang]{Yan2010}
Z.~Yan, J.~Yan, Y.~Hu, and S.~Dang.
\newblock Crystallization in zr60al15ni25 bulk metallic glass subjected to
  rolling at room temperature.
\newblock \emph{Science in China Series E: Technological Sciences}, 53\penalty0
  (1):\penalty0 278--283, Jan 2010.
\newblock ISSN 1869-1900.
\newblock \doi{10.1007/s11431-009-0315-x}.
\newblock URL \url{https://doi.org/10.1007/s11431-009-0315-x}.

\bibitem[Yilmaz(2011)]{Yilmaz2011}
A.~Yilmaz.
\newblock The portevin–le chatelier effect: a review of experimental
  findings.
\newblock \emph{Science and Technology of Advanced Materials}, 12:\penalty0
  063001--1--063001--16, 2011.

\bibitem[Yu et~al.(2015)Yu, Legros, and Minor]{Legros2015}
Q.~Yu, M.~Legros, and A.~M. Minor.
\newblock In situ tem nanomechanics.
\newblock \emph{MRS Bulletin}, 40\penalty0 (1):\penalty0 62–70, 2015.
\newblock \doi{10.1557/mrs.2014.306}.

\bibitem[Yu et~al.(2021{\natexlab{a}})Yu, Chatterjee, Roche, Po, and
  Marian]{YuMarian2021}
Q.~Yu, S.~Chatterjee, K.~J. Roche, G.~Po, and J.~Marian.
\newblock Coupling crystal plasticity and stochastic cluster dynamics models of
  irradiation damage in tungsten.
\newblock \emph{Modelling and Simulation in Materials Science and Engineering},
  29\penalty0 (5):\penalty0 055021, jun 2021{\natexlab{a}}.
\newblock \doi{10.1088/1361-651X/ac01ba}.
\newblock URL \url{https://dx.doi.org/10.1088/1361-651X/ac01ba}.

\bibitem[Yu et~al.(2021{\natexlab{b}})Yu, Martínez, Segurado, and
  Marian]{Yu2021}
Q.~Yu, E.~Martínez, J.~Segurado, and J.~Marian.
\newblock A stochastic solver based on the residence time algorithm for crystal
  plasticity models.
\newblock \emph{Computational Mechanics}, 68:\penalty0 1369--1384,
  2021{\natexlab{b}}.

\bibitem[Zepeda-Ruiz et~al.(2021)Zepeda-Ruiz, Stukowski, Oppelstrup, Bertin,
  Barton, Freitas, and Bulatov]{ZepedaRuiz2021}
L.~A. Zepeda-Ruiz, A.~Stukowski, T.~Oppelstrup, N.~Bertin, N.~R. Barton,
  R.~Freitas, and V.~V. Bulatov.
\newblock Atomistic insights into metal hardening.
\newblock \emph{Nature Materials}, 20:\penalty0 315--320, 2021.

\bibitem[Zhang et~al.(2017)Zhang, Salman, Zhang, Liu, Weiss, Truskinovsky, and
  Sun]{Zhang2017}
P.~Zhang, O.~U. Salman, J.~Y. Zhang, G.~Liu, J.~Weiss, L.~Truskinovsky, and
  J.~Sun.
\newblock Taming intermittent plasticity at small scales.
\newblock \emph{Acta Materialia}, 128:\penalty0 351--364, 2017.

\bibitem[Zhang et~al.(2020)Zhang, Salman, Weiss, and
  Truskinovsky]{ZhangSalman2020}
P.~Zhang, O.~U. Salman, J.~Weiss, and L.~Truskinovsky.
\newblock Variety of scaling behaviors in nanocrystalline plasticity.
\newblock \emph{Phys. Rev. E}, 102:\penalty0 023006, Aug 2020.
\newblock \doi{10.1103/PhysRevE.102.023006}.
\newblock URL \url{https://link.aps.org/doi/10.1103/PhysRevE.102.023006}.

\bibitem[Zhu et~al.(2016)Zhu, To, Ehmann, Xiao, and Zhu]{Zhu2016}
Z.~Zhu, S.~To, K.~F. Ehmann, G.~Xiao, and W.~Zhu.
\newblock A novel diamond micro-/nano-machining process for the generation of
  hierarchical micro-/nano-structures.
\newblock \emph{Journal of Micromechanics and Microengineering}, 26\penalty0
  (3):\penalty0 035009, feb 2016.
\newblock \doi{10.1088/0960-1317/26/3/035009}.
\newblock URL \url{https://dx.doi.org/10.1088/0960-1317/26/3/035009}.

\end{thebibliography}
\restoregeometry

\section*{Appendix A: Finite element solver and time discretisation of the time-discontinuous elasto-plasticity model}

The time interval $\mathcal{T}=[0 \ T]$ is divided into $N$ time steps. The equation of continuum mechanics {\color{black}is enforced} at discrete times $\bar{\mathcal{T}} = \{ t_0 , t_1 , \, ... \, , t_{N} \}$, using the finite element method. In the following, $n$ will be an integer between 0 and $N$. 

\paragraph{\textbf{Finite element solver}}

The domain $\Omega$ is discretised into the set $\mathcal{E}$ of finite elements such that $\Omega \approx  \Omega^h = \bigcup_{e \in \mathcal{E}} \Omega^{(e)} $. {\color{black}The space of linear functions over each element $ \Omega^{(e)}$  is denoted $\mathcal{P}^1 \left( \Omega^{(e)}\right)$.} The finite element space $\mathcal{U}_{n+1}$ in which the displacement field $\vect{u}_{n+1}$ at time $t_{n+1}$ will be looked for is defined by:
\begin{equation}
\mathcal{U}_{n+1} = \left\{ u \in C^0(\Omega) | \forall e \in \mathcal{E}, u_{|\Omega^{(e)}} \in \mathcal{P}^1 \left( \Omega^{(e)} \right) , u = {u_d}_{|t=t_{n+1}} \ \textrm{over} \ \partial_\mathrm{u}   \Omega  \right\}
\end{equation}
and {\color{black}the associated vector space of displacement fields that vanish over $\partial_\mathrm{u}  \Omega$:}
\begin{equation}
\mathcal{U}_0 = \left\{ u \in C^0(\Omega) | \forall e \in \mathcal{E}, u_{|\Omega^{(e)}} \in \mathcal{P}^1 \left( \Omega^{(e)} \right) , u = 0 \ \textrm{over} \ \partial_\mathrm{u}   \Omega  \right\}
\end{equation}

The weak form of the time-discontinuous elasto-plasticity equations at time $t_{n+1} \in \bar{\mathcal{T}} $, discretised by the finite element method, is now introduced. For any virtual displacement $v \in \mathcal{U}_0$, the displacement field $\vect{u}_{n+1} \in \mathcal{U}_{n+1}$ {\color{black}is searched} such that the finite element stress field $ \tenstwo{\sigma}_{n+1}$ at time $t_{n+1}$ satisfies (in absence of volumetric forces and with trivial Neumann boundary conditions):
\begin{equation}
\label{eq:R}
R_{n+1} \left( \vect{v} ; \vect{u}_{n+1} \right) = \int_\Omega \tenstwo{\nabla}_s \vect{v} : \tenstwo{\sigma} \left(  \tenstwo{\nabla}_s  \vect{u}_{n+1} ;  \tenstwo{\varepsilon}^\mathrm{p}_{n}  , p_{n} \right)  \, d\Omega = 0
\end{equation}
where $ \tenstwo{\nabla}_s  \vect{u}_{n+1}  = {\tenstwo{\varepsilon}}_{n+1} $ is the finite element strain field at time $t_{n+1}$. The dependency of the stress on the local history of the strain tensor is introduced via the functional dependency of $ \tenstwo{\sigma}$ on the internal variables ${\tenstwo{\varepsilon}^\mathrm{p}_{n}}  $ and $ p_{n}$  computed at time step $t^n$. 

\paragraph{\textbf{Time discretisation of the constitutive relation}}

The time-discontinuous constitutive equations introduced in the previous section need to be appropriately discretised in time. More precisely, Eq. \eqref{eq:normality2} ,\eqref{eq:yield_cond2}, \eqref{eq:surf2} and \eqref{eq:max} {\color{black}must be replaced} by their time-discrete counterparts so that $\tenstwo{\sigma} \left(  \tenstwo{\nabla}_s  \vect{u}_{n+1} ;  \tenstwo{\varepsilon}^\mathrm{p}_{n}  , p_{n} \right) $ can be evaluated numerically. A fully implicit time-stepping scheme {\color{black}is used}. Eq. \eqref{eq:elast} is enforced at time $t_{n+1}$:
\begin{equation}
 \tenstwo{\sigma}_{n+1} = \tensfour{C} : ( \tenstwo{\varepsilon}_{n+1} -  \tenstwo{\varepsilon}^\mathrm{p}_{n+1} ) 
\end{equation}
{\color{black}The normality rule} Eq. \eqref{eq:normality2} {\color{black}is discretised} as follows:
\begin{equation}
\label{eq:normality3}
\tenstwo{\varepsilon}^\mathrm{p}_{n+1}  - \tenstwo{\varepsilon}^\mathrm{p}_{n} = (p_{n+1} - p_n) \tenstwo{n}^-_{n+1}
\end{equation}
Then, introducing the set $\mathcal{P}_{n+1}$ of cumulative plastic strain increments ${\Delta p^\star}$ satisfying
\begin{equation}
\label{eq:yield_cond3}
 {\Delta p^\star}  f_{n+1}^\star = 0
\end{equation}
and
\begin{equation}
\label{eq:surf3}
 {\Delta p^\star}  \left< \Delta p_\mathrm{min} - {\Delta p^\star} \right> = 0
\end{equation}
the maximum value in set $\mathcal{P}_{n+1}$ {\color{black} is chosen} as the solution of the constitutive update, \textit{i.e.}
\begin{equation}
p_{n+1} = \underset{ \Delta p^\star \in\mathcal{P}_{n+1}  }{ \max  } \Delta p^\star  + p_n 
\end{equation}
In the previous set of equations, $\tenstwo{n}^-_{n+1}$ denotes the normal to the yield surface at time $t_{n+1}$, assuming that no plastic jump took place between $t_n$ and $t_{n+1}$. Mathematically, this means that  $\tenstwo{n}^-_{n+1} = \left. \frac{\partial f }{\partial \sigma} \right|_{ \tenstwo{\sigma}_{n+1} = \tensfour{C} : ( \tenstwo{\varepsilon}_{n+1} -  \tenstwo{\varepsilon}^\mathrm{p}_{n} ) , \, p_{n+1} = p_n} $. Notation $f_{n+1}^\star$ stands for the value of yield function $f$  at the end of the time step, \textit{i.e.} $
f_{n+1}^\star = f( \tensfour{C} : ( \tenstwo{\varepsilon}_{n+1} -  \tenstwo{\varepsilon}^\mathrm{p}_{n}   -  {\Delta p^\star}  \tenstwo{n}^-_{n+1} ; p_n +  {\Delta p^\star} ) $.
Knowing $\tenstwo{\varepsilon}_{n+1}$, the equations of the time-discrete constitutive law are solved algorithmically, and pointwise, to obtain the values of $p_{n+1}$, $\tenstwo{\varepsilon}^\mathrm{p}_{n+1}$, and $\tenstwo{\sigma}_{n+1}$.
This local algorithmic  procedure (the so-called constitutive update) will be detailed in the next section.

\paragraph{\textbf{Global Newton solver}}

The Newton-Raphson solver used to solve the nonlinear finite element system of equations at time $t_{n+1}$ requires the linearisation of residual \eqref{eq:R}. The Gateau-derivative of $R_{n+1}$ in direction $\delta u \in \mathcal{U}_0$ is the following bilinear form (linear both in $\vect{\delta u}$ and in $ \vect{v}$):
\begin{equation}
\delta R_{n+1} \left( \vect{\delta u} , \vect{v} ; \vect{u}_{n+1} \right) = \int_\Omega \tenstwo{\nabla}_s \vect{v} : \left. \frac{ \partial \tenstwo{\sigma}_{n+1} }{\partial  \tenstwo{\varepsilon}_{n+1}  } \right|_{\tenstwo{\varepsilon}_{n+1} = \tenstwo{\nabla}_s  \vect{u}_{n+1} } : \tenstwo{\nabla}_s  \vect{\delta u}  \, d\Omega
\end{equation}

The discrete equations of time-discontinuous elasto-plasticity may be formally solved using a Newton algorithm. At iteration $k$ of the Newton procedure, we look for a finite element displacement correction $\vect{\Delta u}_{n+1,k+1} =  \vect{u}_{n+1,k+1}  -  \vect{u}_{n+1,k} \in \mathcal{U}_0$ that satisfies
\begin{equation}
\delta R_{n+1} \left( \vect{\Delta u}_{n+1,k+1} , \vect{v} ; \vect{u}_{n+1,k} \right) = - R_{n+1} \left( \vect{v} ; \vect{u}_{n+1,k} \right) 
\end{equation}
for all virtual displacements $\vect{v} \in \mathcal{U}_0$. Newton initialisation $ \vect{u}_{n+1,0}$ must be chosen such that it satisfies the Dirichlet conditions at time $t_{n+1}$, i.e. $\vect{u}_{n+1,0} \in \mathcal{U}_{n+1}$. 

When solving the finite element system of equations at time $t_{n+1}$, the stress field $\tenstwo{\sigma}_{n+1} $ needs to be evaluated at the quadrature points of the finite element mesh. Internal variables from $t_n$, namely $\tenstwo{\varepsilon}^\mathrm{p}_{n} $ and $ p_{n}$, are stored at quadrature points, following the algorithmic procedure detailed {\color{black} in section \ref{sec:ConstitutiveUpdate}}.

\end{document}